\documentclass[letter,oneside,9pt,twocolumn,article]{aiaa}
\usepackage{graphicx}
\usepackage{rotating}
\usepackage{bm}
\usepackage{flushend}
\usepackage{cite}
\usepackage[bf,normalsize]{subfigure}
\usepackage{adjustbox}
\usepackage{tabularx}
\usepackage{multirow}
\usepackage{multicol}
\usepackage{indentfirst} 
\usepackage[hidelinks]{hyperref}
\usepackage{float}
\usepackage{xcolor}
\usepackage[round]{natbib}
\setcitestyle{authoryear,open={(},close={)}} 

\usepackage{mwe}
\usepackage[markcase=noupper]{scrlayer-scrpage}
\usepackage{caption} 

\ihead*{\href{https://doi.org/10.1063/5.0244163}{{\it Physics of Fluids}, 37, 013609 (2025)}}
\ohead*{\pagemark}
\ifoot*{\rm }
\ofoot*{\rm Unrestricted content}
\newcommand{\alb}{\vspace{0.1cm}\\} 
\newcommand{\mfd}{\displaystyle}
\renewcommand{\vec}[1]{\bm{#1}}

\renewcommand{\fontsizetable}{\footnotesize\scalefont{0.9}}

\renewcommand{\vec}[1]{\bm{#1}}
\author{Felipe Martin Rodriguez Fuentes\thanks{Graduate Student}~~and~
Bernard Parent\thanks{Associate Professor, bparent@arizona.edu.}\\[0.3em] \it University of Arizona, Tucson, AZ 85721, USA.
}

\title{Impact of Ion Mobility on Electron Density\\ and Temperature in Hypersonic Flows}

\abstract{ 
This study provides the first comprehensive analysis of how ion mobility affects electron density and temperature in hypersonic flows. We compare two ion mobility models: one derived from Gupta-Yos cross-sections, and the other from swarm drift velocity experiments. The ion mobility model significantly alters the plasma density around a hypersonic waverider, with increases of more than twofold observed at low dynamic pressures and high Mach numbers. This is partly due to electron loss through surface catalysis, which depends on ambipolar diffusion scaling with ion mobility. We also derive novel scaling laws that highlight the strong dependence of electron cooling on ion mobility both within the quasi-neutral regions and the non-neutral plasma sheaths. Electron cooling influences the electron temperature across the plasma, leading to a previously unrecognized impact of ion mobility on plasma bulk temperature. This in turn affects plasma density via electron-ion recombination rates which are temperature-dependent. Accurately modeling ion mobility is critical for predicting hypersonic plasma behavior, with important implications for optimizing magnetohydrodynamic technologies and mitigating or exploiting plasma-induced interference with electromagnetic waves.
}

\nomenclature{
\begin{nomenclaturelist}{Roman Symbols}
{
\item[$C_k$]{charge of $k$th  species, C}
\item[$(c_p)_k$]{specific heat at constant pressure of $k$th species, $\rm J/(kg\cdot K)$}
\item[$d_{21}$]{distance between stations 1 and 2, m}
\item[$d_{\rm pw}$]{distance between peak plasma location and wall, m}
\item[$d_{\rm ps}$]{distance between peak plasma location and shock, m}
\item[$D_{\rm a}$]{ambipolar diffusion coefficient, $\rm m^2/s$}
\item[$E_{\rm i}$]{electric field along $i$th dimension, V/m}
\item[$\vec{E}$]{electric field vector, V/m}
\item[$E^\star$]{reduced electric field, $E/N$, $\rm Vm^2$}
\item[$\mathcal{E}_{kl}$]{ activation energy of the $l$th electron impact process of the $k$th species, J/particle}
\item[$e_{\rm e}$]{electron specific internal energy, J/kg}
\item[$e_{\rm t}$]{total specific internal energy, J/kg}
\item[$e_{\rm v}$]{nitrogen vibration specific energy, J/kg}
\item[$h_{k}^{0}$]{heat of formation of $k$th species, J/kg}
\item[$h_k$]{specific enthalpy of $k$th species excluding heat of formation, J/kg}
\item[$h_{\rm e}$]{electron specific enthalpy, J/kg}
\item[$\vec{J}$]{current density vector, A/m$^2$}
\item[$\vec{J}_{\rm e}$]{electron current density vector, A/m$^2$}
\item[$k_{\rm B}$]{Boltzmann constant, J/K}
\item[$k_{kl}$]{electron impact reaction rate of $l$th process on $k$th neutral species, $\rm m^3/s$}
\item[$\rm Kn$]{Knudsen number}
\item[$L_{\rm sheath}$]{plasma sheath length scale, m}
\item[${\cal M}_i$]{molecular weight of $i$th species, g/g-mol}
\item[$\rm M$]{Mach number}
\item[$m_k$]{particle mass of $k$th species, kg}
\item[$m_{\rm i}$]{ion mass, kg}
\item[$N_k$]{number density of $k$th species, $\rm m^{-3}$}
\item[$N_{\rm e}$]{number density of electrons, $\rm m^{-3}$}
\item[$N_{\rm i}$]{number density of ions, $\rm m^{-3}$}
\item[$N_{\rm n}$]{number density of neutral species, $\rm m^{-3}$}
\item[$N$]{number density of gas mixture, $\rm m^{-3}$ or $\rm cm^{-3}$}
\item[$\vec{n}$]{normal vector to surface}
\item[$P_k$]{partial pressure of $k$th species, Pa}
\item[$P_{\rm e}$]{electron pressure, atm}
\item[$P$]{pressure of the gas mixture, $\sum P_k$, Pa}
\item[$P_{\rm dyn}$]{flight dynamic pressure, Pa}
\item[$Q_{\rm J}^{\rm e} $]{electron Joule heating rate, W/m$^3$ }
\item[$Q_{\rm e} $]{electron energy relaxation due to elastic and inelastic collisions, W/m$^3$ }
\item[$Q_{\rm elastic} $]{electron heating-cooling rate due to elastic collisions, W/m$^3$ }
\item[$Q_{\rm inelastic} $]{electron cooling rate due to inelastic collisions, W/m$^3$ }
\item[$q$]{magnitude of velocity vector, m/s}
\item[$q_i$]{heat flux component along $i$th dimension, $\rm W/m^2$}
\item[$\overline{{q}_{\rm i}}$]{ion thermal speed, m/s}
\item[$R$]{leading edge radius, m}
\item[$\mathcal{R}$]{universal gas constant, $\rm cal/(g-mol \cdot K)$}
\item[$s_k$]{species sign according to charge, $\pm 1$}
\item[$S$]{surface area, $\rm m^2$}
\item[$T$]{bulk gas temperature, K}
\item[$T_{\rm ref}$]{reference temperature in experiments, K}
\item[$T_{\rm v}$]{vibrational temperature, K}
\item[$T_{\rm e}$]{electron temperature, K}
\item[$t$]{time, s}
\item[$U$]{velocity magnitude on streamline, m/s}
\item[$\vec{V}_{\rm e}$]{electron velocity vector, m/s}
\item[$\vec{V}$]{bulk velocity vector, m/s}
\item[$v_{\rm dr}$]{ion drift velocity, m/s}
\item[$V_{i}$]{bulk flow velocity component, m/s}
\item[$V_{i}^k$]{component of $k$th species velocity, m/s}
\item[$V$]{volume, $\rm m^3$}
\item[$W_{\rm e}$]{chemical source term of electrons, $\rm kg/(m^3\cdot s)$ }
\item[$W_k$]{chemical source term of $k$th species, $\rm kg/(m^3\cdot s)$ }
\item[$w_k$]{mass fraction of $k$th species}
\item[$w_{\rm e}$]{electron mass fraction}
\item[$w_{\rm N_2}$]{nitrogen mass fraction}
\item[$x_{i}$]{Cartesian coordinate, m}
}

\end{nomenclaturelist}

\begin{nomenclaturelist}{Greek Symbols}
{
\item[$\beta_{k}^{\rm n}$]{parameter equal to 1 when $k$th species is neutral, 0 otherwise}
\item[$\Delta_{ij}^{(1)}$]{collision parameter, $\rm cm \cdot s$}
\item[$\epsilon_0$]{ permittivity of free space, $\rm C/(V\cdot m)$}
\item[$\eta$]{ non-dimensional ambipolar diffusion ratio}
\item[$\vec{\Gamma}$]{number flux, $\rm particles/(m^{2}\cdot s)$}
\item[$\gamma_{\rm e}$]{electron secondary emission coefficient}
\item[$\kappa$]{thermal conductivity, $\rm W/(m\cdot K)$}
\item[$\kappa_{\rm e}$]{electron thermal conductivity, $\rm W/(m\cdot K)$}
\item[$\kappa_{\rm i}$]{ion thermal conductivity, $\rm W/(m\cdot K)$}
\item[$\kappa_{\rm n}$]{neutrals thermal conductivity, $\rm W/(m\cdot K)$}
\item[$\kappa_{\rm react}$]{reactive  thermal conductivity, $\rm W/(m\cdot K)$}
\item[$\kappa_{\rm total}$]{total thermal conductivity, $\rm W/(m\cdot K)$}
\item[$\kappa_{\rm v}$]{$\rm N_2$ vibrational thermal conductivity, $\rm W/(m\cdot K)$}
\item[$\lambda_{\rm D}$]{Debye length, m}
\item[$\ln \Lambda$]{Coulomb logarithm}
\item[$\mu_{\rm ii}$]{mobility in ion-ion collisions, $\rm m^2/(V\cdot s)$}
\item[$\mu_{\rm in}$]{mobility in ion-neutral collisions, $\rm m^2/(V\cdot s)$}
\item[$\mu_{\rm i}$]{ion mobility, $\rm m^2/(V\cdot s)$}
\item[$\mu_k$]{mobility of $k$th charged species,$\rm m^2/(V\cdot s)$}
\item[$\mu_{\rm e}^\star$]{reduced electron mobility, $\mu_{\rm e}N$, $\rm m^{-1}V^{-1}s^{-1}$}
\item[$\Omega_{ij}^{(1,1)}$]{average cross-section of $ij$ collision pair, $\r{A}^2$}
\item[$\rho_{k}$]{partial mass density of $k$th species, kg/m$^3$}
\item[$\rho_c$]{net charge density, C/m$^3$}
\item[$\rho_{\rm e}$]{electron mass density, kg/m$^3$}
\item[$\rho_{\rm N_2}$]{nitrogen mass density, kg/m$^3$}
\item[$\rho$]{mass density of the mixture, $\sum_k \rho_k$, kg/m$^3$}

\item[$\sigma_{\rm in}$]{ion-neutral collision cross section, m$^2$}
\item[$\tau_{ji}$]{viscous stress tensor, Pa}
\item[$\tau_{\rm vt}$]{vibration-relaxation time, s}
\item[$\nu_k$]{mass diffusion coefficient of $k$th species, $\rm kg/(m\cdot s)$}
\item[$\nu_{\rm N_2}$]{mass diffusion coefficient of nitrogen, $\rm kg/(m\cdot s)$}
\item[$\nu_{\rm ii}$]{frequency of ion-ion collisions, 1/s}
\item[$\nu_{\rm in}$]{frequency of ion-neutral collisions, 1/s}
\item[$\chi$]{wall-normal coordinate, m}
\item[$\chi_k$]{mole fraction of $k$th species}
\item[$\xi$]{non-dimensional constant in collision frequency}
\item[$\zeta_{\rm v}$]{Joule heating fraction to vibrational energy levels}
}

\end{nomenclaturelist}

}

\begin{document}
\maketitle
\makenomenclature

\section{Introduction}
\dropword 
Plasma~layers that form around hypersonic and reentry vehicles present a range of challenges. One longstanding issue is the communication blackout, which occurs when the plasma density is high enough to attenuate electromagnetic waves, thereby disrupting radio communications [\cite{ieee:1971:rybak, aiaaconf:2007:hartunian,aiaaconf:2009:kim, aiaaconf:2011:davis}]. Additionally, the plasma layer extends into the vehicle's wake, spanning tens or hundreds of meters. This large volume of low-density wake plasma can be effectively detected with radar {[\cite{aiaa:1964:lees, phdthesis:2012:fenstermacher,ieee:2019:wang, jst:2024:esposito}]}. Conversely, the plasma layer can also be utilized to conceal the vehicle by absorbing electromagnetic waves with frequencies originating from radio to laser sources [\cite{gmrl:1963:musal, dtic:1992:gregorie, spie:2020:xu}]. 

Detailed knowledge of the variation of the electron density, collision frequency, and temperature in the plasma layer would lead to a better evaluation of these various interference phenomena. Such insights could lead to the development of new technologies aimed at mitigating or harnessing the interference caused by plasma layers on electromagnetic waves.

Beyond interference with electromagnetic waves, plasma density significantly impacts other hypersonic technologies, including magnetohydrodynamic (MHD) force and power generators. Studies by \cite{aiaaconf:2022:moses} and \cite{jtht:2023:parent, aiaaconf:2024:parent} have demonstrated that the interaction between a high-speed plasma flow and a small magnetic field of 0.1–0.2 Tesla generates substantial Lorentz forces around planetary entry vehicles. These forces can alter the vehicle’s trajectory by creating additional lift and drag or be harnessed to recharge batteries and power on-board devices by extracting energy from the flow. Since Lorentz forces scale with the plasma's electrical conductivity -- and electrical conductivity in a weakly-ionized plasma increases proportionally with electron molar fraction -- the performance of MHD devices is highly sensitive to changes in plasma density.

Previous efforts to accurately determine electron density have primarily focused on air chemical kinetics.  \cite{agard:1964:eschenroeder} outlined three critical processes: the dissociation of molecular nitrogen and oxygen into their atomic components; the associative ionization of nitric oxide from atomic oxygen and nitrogen; and electron-ion recombination. Early models by \cite{misc:1964:lenard} and \cite{nasa:1973:dunn} relied solely on translational temperature as the controlling parameter for reaction rates. Subsequently, two-temperature models were developed to account for vibrational temperature in non-equilibrium conditions, the \cite{book:1990:park} model being the most widely adopted. Despite various enhancements to the Park model (e.g.\ \cite{aiaa:202:chaudhry, ijhmt:2021:kim, jtht:2024:aitken}), these models have not resulted in changes exceeding a twofold variation in plasma density.

To date, no studies have isolated the effect of ion mobility (or the diffusion coefficient of ions which is related to their mobility) on the final plasma properties. Typically, ion mobilities in hypersonic flows are evaluated using the \cite{nasa:1990:gupta} model alongside other transport coefficients {although there are a few exceptions such as the recent work by \cite{psst:2024:petrova}.}
Ion mobility is crucial for determining electron density because, for instance, the electron flux due to ambipolar diffusion depends not only on electron temperature but also on the ion mobility. Ambipolar diffusion can account for a large portion of electron losses as reported in \cite{pf:2022:parent}, thereby directly affecting plasma density.

In this paper, we aim to assess the extent to which ion mobility affects the properties of the plasma layer surrounding a hypersonic vehicle. We begin by outlining the physical model, including descriptions of the two ion mobility models considered: the \cite{nasa:1990:gupta} model and one derived from drift velocities obtained from swarm experiments by \cite{book:1997:grigoriev} and \cite{misc:1968:sinnott}. This is followed by a detailed evaluation of the impact of ion mobility on electron density and temperature surrounding both a generic hypersonic waverider and the RAM-C-II re-entry vehicle.

\section{Physical Model}

As noted in \cite{pf:2021:parent},  hypersonic plasma flows can be significantly influenced by the non-neutral sheaths that form  near the surfaces of the vehicle.  The motion of electrons and ions relative to the bulk thus cannot be approximated using the ambipolar diffusion coefficient because such is only a valid approximation in the quasi-neutral regions. The velocities of the charged species are rather here {determined by the drift-diffusion model described in \cite{book:2022:parent}} because such is appropriate for both non-neutral sheaths and quasi-neutral bulk regions. On the other hand, the velocities of the neutral species are obtained from Fick's law of diffusion. Thus, the $i$th component of the $k$th species velocity (whether charged or neutral) can be expressed as:
\begin{equation}
  V^{k}_i = \left\{
  \begin{array}{ll}\mfd
  V_i+s_k \mu_k  {E}_i
             -  \frac{\mu_k}{|C_k| N_k} \frac{\partial P_k}{\partial x_i} & \textrm{for electrons and ions} \alb\mfd
  V_i - \frac{\nu_k}{\rho_k} \frac{\partial w_k}{\partial x_i} & \textrm{for neutrals}
  \end{array}
  \right.
\label{eqn:Vk}
\end{equation}
where $\mu_k$ is the mobility of the $k$th species, $N_k$ is the species number density, $E_i$ is the $i$th component of the electric field vector, $s_k$ is the species sign (either +1 or -1), $C_k$ is the species charge in Coulombs, $\nu_k$ is the mass diffusion coefficient obtained from the \cite{nasa:1990:gupta} model,  and $w_k$ is the species mass fraction.

The mass conservation equation for either the neutral or charged species as well as the momentum conservation equation for the bulk {(Navier-Stokes equations)} take on the following form:
\begin{equation}
\frac{\partial}{\partial t} \rho_k + \sum_i \frac{\partial }{\partial x_i}\rho_{k} V_i^{k} = W_{k}  
\label{eqn:masstransport}
\end{equation}
\begin{equation}
  \rho \frac{\partial V_i }{\partial t}+ \sum_{j=1}^3 \rho V_j \frac{\partial V_i}{\partial x_j}
=
-\frac{\partial P}{\partial x_i} 
+ \sum_{j=1}^3 \frac{\partial \tau_{ji}}{\partial x_j}
+ \rho_{\rm c}{E}_i
\label{eqn:momentum}
\end{equation}
In the latter $\rho_k$ is the species mass density, $P_k$ the species partial pressure, while $\rho_{\rm c}$ is the net charged density, and $\vec{V}$ is the bulk flow velocity vector. Further, $W_k$ are the chemical source terms of the $k$th species with the chemical reactions and rates obtained from {\cite{pf:2022:parent}} but with the reactions involving electrons in the reactants given reaction rates outlined in \cite[Table 4]{pf:2024:parent}. Also, $P$ is the pressure of the mixture corresponding to the sum of the partial pressures while $\tau_{ij}$ is the viscous stress tensor with the viscosity obtained from the \cite{nasa:1990:gupta} model. 

An energy transport equation for the nitrogen vibrational temperature {derived from the first law of thermodynamics} is also solved:
\begin{equation}
 \begin{array}{l}
  \mfd\frac{\partial}{\partial t} \rho_{\rm N_2} e_{\rm v}
     + \sum_{j=1}^{3} \frac{\partial }{\partial x_j}
       \rho_{\rm N_2} V_j e_{\rm v}
     - \sum_{j=1}^{3} \frac{\partial }{\partial x_j} \left(
            \kappa_{\rm v}  \frac{\partial T_{\rm v}}{\partial x_j}\right)\alb\mfd
     - \sum_{j=1}^{3} \frac{\partial }{\partial x_j} \left(
            e_{\rm v} \nu_{\rm N_2}  \frac{\partial w_{\rm N_2}}{\partial x_j}\right)
 = 
  \frac{N_{\rm N_2}}{N} \zeta_{\rm v} Q_{\rm J}^{\rm e}   + {\frac{\rho_{\rm N_2}}{\tau_{\rm vt}}}\left( e_{\rm v}^0 -e_{\rm v} \right) + W_{\rm N_2} e_{\rm v}
\end{array}
\label{eqn:vibrationalenergy}
\end{equation}
where $e_{\rm v}$ is the nitrogen vibrational energy, $T_{\rm v}$ is the nitrogen vibrational energy, $\tau_{\rm vt}$ is the vibration-relaxation time taken from \cite{aiaa:2001:macheret,aiaaconf:1999:macheret}. Also, $\zeta_{\rm v}$ is the Joule heating fraction that is consumed by exciting the nitrogen vibrational energy levels taken from \cite{pf:2024:parent}, and  $Q_{\rm J}^{\rm e}$ is the electron Joule heating taken from the Appendix of \cite{aiaa:2016:parent}.  

The gas temperature $T$ is obtained from the total energy transport equation outlined in {\cite{pf:2021:parent}}:
\begin{equation}
\begin{array}{l}\mfd
 \frac{\partial }{\partial t}\rho e_{\rm t}
+ \sum_{j=1}^3  \frac{\partial }{\partial x_j} V_j \left(\rho  e_{\rm t} +  P \right)
 \alb\mfd
- \sum_{j=1}^3  \frac{\partial }{\partial x_j} \left(
   \nu_{\rm N_2} e_{\rm v}\frac{\partial w_{\rm N_2}}{\partial x_j} 
  + \sum_{k=1}^{n_{\rm s}}  \rho_k (V^k_j-V_j) {(h_k+h_k^\circ)}  
\right)
 \alb\mfd
-\sum_{i=1}^{3}\frac{\partial }{\partial x_i}\left((\kappa_{\rm n}+\kappa_{\rm i}) \frac{\partial T}{\partial x_i} 
+ \kappa_{\rm v} \frac{\partial T_{\rm v}}{\partial x_i} +\kappa_{\rm e} \frac{\partial T_{\rm e}}{\partial x_i}\right)
 \alb\mfd
=
 \sum_{i=1}^3 \sum_{j=1}^3  \frac{\partial }{\partial x_j} \tau_{ji} V_i
+ \vec{E}\cdot\vec{J}
\end{array}
\label{eqn:totalenergy}
\end{equation}
where $e_{\rm t}$ is the total specific energy of the mixture, $h_k$ the species specific enthalpy, $h_k^\circ$ the species heat of formation, $\kappa_{\rm i}+\kappa_{\rm n}$ the sum of the thermal conductivities of ions and neutrals, and $\vec{J}$ the current density vector. The sum of the specific enthalpy and heat of formation is taken from \cite{nasa:2002:mcbride}. 

Furthermore, the electron temperature $T_{\rm e}$ is obtained from the following transport equation, {as found in \cite{pf:2024:parent} or in \cite{book:1991:raizer}}:
\begin{equation}
 \frac{\partial }{\partial t}\rho_{\rm e} e_{\rm e} + \sum_{j=1}^3  \frac{\partial }{\partial x_j} \left( \rho_{\rm e}{V}^{\rm e}_j h_{\rm e}  - \kappa_{\rm e} \frac{\partial T_{\rm e}}{\partial x_j}  
\right)
= 
 W_{\rm e} e_{\rm e}
+  \vec{E} \cdot \vec{J}_{\rm e}  
- Q_{\rm e} 
    \label{eqn:ee_transport}
\end{equation}
with $\vec{J}_{\rm e}$ the electron current,  $\kappa_{\rm e}$ the electron thermal conductivity,  and $e_{\rm e}$ the electron specific internal energy. Also the electron cooling-heating term $Q_{\rm e}$ corresponds to the sum of electron cooling due to elastic and inelastic collisions, i.e.\ $Q_{\rm e}=Q_{\rm elastic}+ Q_{\rm inelastic}$ with the electron cooling{-heating} from elastic collisions taken from \cite{pf:2024:parent}. 
Further, as shown in \cite{pf:2024:parent}, the electron cooling due to inelastic collisions from all processes can be expressed without any loss in generality as a function of the reduced electron mobility and reduced electric field as follows:
\begin{equation}
  \sum_l k_{kl} \mathcal{E}_{kl}  
=  |C_{\rm e}|  \left(\mu_{\rm e}^\star\right)_k \left( (E^\star_k)^2 - \frac{3  k_{\rm B}    (T_{\rm e}-T_{\rm ref})}{ m_{k} (\mu_{\rm e}^\star)^2_k}\right) 
\end{equation}
where $l$ denotes an electron impact process, $k_{\rm B}$ is the Boltzmann constant, $k_{kl}$ is the rate coefficient of the $l$th electron impact process acting on the $k$th neutral species, $\mathcal{E}_{kl}$ is the activation energy of the $l$th electron impact process of the $k$th species. Also, $\mu_{\rm e}^\star=\mu_{\rm e} N$ (with $N$ the total number density of the mixture) is the reduced electron mobility while $E^\star=|\vec{E}|/N$ is the reduced electric field, and $T_{\rm ref}$ is the gas temperature of the swarm experiments needed to find the reduced mobility and reduced electric field as a function of electron temperature. {Spline curve fits of the reduced electric field and the reduced mobility for all neutral species are taken from \cite{pf:2024:parent}.}

Following \cite{jap:2010:shneider} and \cite{pf:2021:parent} we here assume that inelastic electron heating uses the same rates as inelastic electron cooling such that the net rate of electron cooling and heating scales with $T_{\rm e}-T_{\rm v}$. We thus obtain the following expression for electron heating and cooling through inelastic collisions:  
\begin{equation}
Q_{\rm inelastic}  
=  \sum_k \beta_k^{\rm n} |C_{\rm e}| N_{\rm e} N_k (\mu_{\rm e}^\star)_k \left( (E^\star_k)^2 -  \frac{3  k_{\rm B}    (T_{\rm e}-T_{\rm ref})}{ m_{k} (\mu_{\rm e}^\star)^2_k} \right)\left(\dfrac{T_{\rm e}-T_{\rm v}}{T_{\rm e}}\right)
\label{eqn:Qinelastic}
\end{equation}
where $\beta_k^{\rm n}$ is equal to 1 when the $k$th species is a neutral and equal to 0 otherwise, and $m_k$ is the mass of one particle.

To close the system of equations, the electric field is here obtained as the negative of the gradient of the electric potential determined from Gauss' law. Boundary conditions assume no surface catalysis for any neutral species. For charged species we consider surface catalysis through the boundary condition outlined in \cite{pf:2024:parent} with the secondary electron emission coefficient set to 0.1.  

Excluding ion mobility, all transport coefficients such as viscosity, thermal conductivity, diffusion coefficients and electron mobility are determined following \cite{nasa:1990:gupta}, {with some modifications outlined in \cite{jtht:2023:parent}. The first modification is to determine the thermal conductivity of the charged species as:
\begin{equation}
  \kappa_k = \frac{\rho_k k_{\rm B} (c_p)_k T_k \mu_k}{|C_k|}    
  \label{eqn:kappak_from_muk}
\end{equation}
The second modification is to exclude electron-electron collisions when determining the electron mobility, since these collisions do not result in loss of momentum of electrons and hence do not affect electron mobility.
}

To determine the mobility of ions, we will consider two approaches. The first approach makes use of cross-sectional data as given in \cite{nasa:1990:gupta}. The second is based on swarm experiments performed by \cite{misc:1968:sinnott} and \cite{book:1997:grigoriev}.

\subsection{Ion Mobilities from Cross-sectional Data (GY model)}

The first model for the ion mobilities considered is one based on collision cross-sections as outlined in \cite{nasa:1990:gupta}:
\begin{equation}
 \mu_i=  |C_i| \left( N k_{\rm B} T \sum_{j=1}^{n_{\rm s}} \chi_j  \Delta_{ij}^{(1)}(T) \right)^{-1} 
\end{equation}
where  $\chi_i$ the mole fraction of the $i$th species and $k_{\rm B}$ the Boltzmann constant in m$^2$kg/s$^2$-K. Also  $N$ is the number density of the plasma in $1/{\rm cm}^3$ and $\Delta_{ij}^{(1)}$ has units of cm$\cdot$s with the following expression,
\begin{equation}
    \Delta_{ij}^{(1)}(T)=(\ln \Lambda)_{ij}\frac{8}{3} (1.5460 \times 10^{-20}) \pi \Omega_{ij}^{(1,1)}(T) \sqrt{\frac{2 {\cal M}_i {\cal M}_j}{\pi {\cal R} T ({\cal M}_i + {\cal M}_j)} }
\end{equation}
In the latter, $\Omega_{ij}^{(1,1)}$ is the average collision cross-section for collisions between species $i$ and $j$ in square Angstroms and $\mathcal{M}_{i}$ the molecular weight of the $i{\rm th}$ species {in g/g-mol}. Expressions and data for $\Omega_{ij}^{(1,1)}(T)$ as a function of temperature can be found in \cite{nasa:1990:gupta}. The Coulomb logarithm $(\ln \Lambda)_{ij}$ is the electron pressure correction which is set here to:
\begin{align}
  &(\ln \Lambda)_{ij}= \nonumber\\
  &\left\{ \begin{array}{ll}\mfd\frac{1}{2} \ln \left( \mfd 0.0209 \left( \frac{T_{\rm e}^4}{1000 P_{\rm e} }\right) + 1.52 \left( \frac{T_{\rm e}^4}{1000 P_{\rm e} }\right)^\frac{2}{3}
  \right)
  & \textrm{\begin{minipage}{2.5cm}\flushleft if species $i$ and $j$ are both charged\end{minipage}}\\
  1 & {\rm otherwise}  
  \end{array}
  \right.
\end{align}
%
%
with $P_{\rm e}$ the electron pressure in atm, $\cal R$ the universal gas constant in cal/(g-mol K), $T_{\rm e}$ the electron temperature in K.

\subsection{Ion Mobilities from Swarm Experiments (SE model)}

The second ion mobility model here considered is not obtained from cross-sectional data but rather from swarm experiments. Because the swarm experiments are performed in weakly-ionized plasmas where ion-ion collisions are negligible and because ion-ion collisions could be important in hypersonic flows, the mobilities need to be corrected as follows: 
\begin{equation}
\mu_{\rm i}=\dfrac{1}{\dfrac{1}{\mu_{\rm in}}+\dfrac{1}{\mu_{\rm ii}}}
\end{equation}
where $\mu_{\rm ii}$ is the ion mobility from ion-ion collisions and $\mu_{\rm in}$ is the ion mobility in collisions of ions with neutral species as obtained from the swarm experiments. We first obtain the former from the ion-ion collision frequency as follows:
\begin{equation}
 \mu_{\rm ii} = \frac{|C_i|}{m_i \nu_{\rm ii}}
   \label{eqn:muii_eqn}
\end{equation}
where $C_i$ is the charge of the ion in Coulomb and $m_{\rm i}$ is the mass of one ion particle in kg. A theoretical expression for the ion-ion collision frequency $\nu_{\rm ii}$ is derived in  \cite{book:1984:chen}:
\begin{equation}
\nu_{\rm ii}=\xi \frac{ N_{\rm i}}{ \sqrt{m_{\rm i}} (k_{\rm B} T_{\rm i})^\frac{3}{2}}  \frac{C_{\rm i}^4}{16 \pi^2 \epsilon_0^2}  \ln \Lambda
  \label{eqn:nuii_eqn}
\end{equation}
{After substituting the latter in the former we obtain:
\begin{equation}
 \mu_{\rm ii} = \frac{16 \pi^2 \epsilon_0^2 (k_{\rm B} T_{\rm i})^\frac{3}{2}}{\mfd \sqrt{m_{\rm i}} \xi  N_{\rm i}  |C_{\rm i}|^3  \ln \Lambda}
\end{equation}
with $\xi$ a non-dimensional parameter  function only of the masses of the colliding ions and varying between  1.71 to 2.36 \cite[page 60]{book:2001:hutchinson}. As well, $\ln \Lambda$ is the Coulomb logarithm listed in the Navy Research Laboratory (NRL) Formulary \cite[page 35]{nrl:2002:huba} for mixed ion-ion collisions, which can be simplified to:
\begin{equation}
\ln\Lambda = 23-\ln\left(T^{-1.5}{N_{\rm i}}^{0.5}\right)
\end{equation}
with the bulk temperature $T$ in eV and the ion number density $N_{\rm i}$ in cm$^{-3}$. For the flowfields under consideration, the Coulomb logarithm typically has a range of 5-7.
We here assign a $\xi\ln\Lambda$ value of 10.8. Such is representative of weakly-ionized high-altitude flows (ion density in the order of $10^{16}~\rm m^{-3}$) where $\xi\ln\Lambda$ is in the range $8.6-16.5$.
 Evaluating all constants, the ion-ion mobility can be written as:}
\begin{equation}
 \mu_{\rm ii} = 14.3 m_{\rm i}^{-0.5} T_{\rm i}^{1.5} N_{\rm i}^{-1}
  \label{eqn:muii}
\end{equation}
with $T_{\rm i}$ the ion translational temperature in Kelvin and $N_{\rm i}$ the total ion number density in $\rm m^{-3}$. We can find the ion mobility in ion-neutral collisions $\mu_{\rm in}$ similarly through the collision frequency between ions and neutrals, following \cite{book:1984:chen} again:
\begin{equation}
    \nu_{\rm in}= N_{\rm n} \overline{q_{\rm i}} \sigma_{\rm in}
\end{equation}
where $N_{\rm n}$ is is the sum of the neutral species number densities and $\sigma_{\rm in}$ is the cross section of collisions of ions with neutrals. The velocity with which an ion approaches a neutral species before a collision is here taken as the ion thermal speed:
\begin{equation}
    \overline{q_{\rm i}} = \sqrt{\frac{8 k_{\rm B} T}{\pi m_{\rm i}}}
\end{equation}
substituting the latter in the former, evaluating all constants, and for single-charge ions, the mobility of ions from ion-neutral collisions becomes:
\begin{equation}
 \mu_{{\rm in}} = N_{\rm n}^{-1}\cdot 2.7021\cdot 10^{-8} (m_{\rm i}T)^{-0.5}\sigma_{{\rm in}}^{-1}   
 \label{eqn:muin}
\end{equation}
In the latter, the dependence of the collision cross section on temperature is small in the case of collisions between heavy particles at high temperatures. Swarm experiments by \cite{misc:1968:sinnott} and \cite{book:1997:grigoriev} report direct measurements of the drift velocity of $\rm N_{2}^+$, $\rm O_{2}^+$ and $\rm NO^+$ ions, $v_{\rm dr}$, as a function of the reduced electric field $E^\star=E/N$. Thus, from the latter, the ion mobility can be obtained:
\begin{equation}
 \mu_{{\rm in}} = \dfrac{v_{\rm dr}}{E^\star N}  
\end{equation}
with $N$ the number of gas molecules per unit volume found from the conditions of the experiments. With the known values of ion mobility, we can obtain the final expressions using Eqs.\ (\ref{eqn:muii}) and (\ref{eqn:muin}). The mobilities of $\rm N^+$ and $\rm O^+$ are assumed to follow the same expressions as $\rm N_{2}^+$, $\rm O_{2}^+$ and $\rm NO^+$ ions but using instead the corresponding ion masses.  The final expressions of the mobilities of positive ions in air are listed in Table~\ref{tab:pm:mu}.

\begin{table}[!t]
  \center\fontsizetable
  \begin{threeparttable}
    \tablecaption{Ion mobilities in dry air obtained from swarm experiments (SE model).\tnote{a,b}}
    \label{tab:pm:mu}
    \fontsizetable
    \begin{tabular*}{\columnwidth}{c@{\extracolsep{\fill}}c}
    \toprule
    Ion & Mobility, $\rm m^2\cdot V^{-1}\cdot s^{-1}$ \\
    \midrule

    N$_2^+$         & $\frac{1}{2}{\rm avgh}\left( N_{\rm n}^{-1} \cdot 0.75\cdot 10^{23}\cdot T^{-0.5}~,~~14.3 \cdot m_{\rm N_2^+}^{-0.5} \cdot T^{1.5} \cdot N_{\rm i}^{-1}\right)$  \alb
    O$_2^+$         &  $ \frac{1}{2}{\rm avgh} \left(N_{\rm n}^{-1} \cdot 1.18\cdot 10^{23}\cdot T^{-0.5},~~14.3 \cdot m_{\rm O_2^+}^{-0.5} \cdot T^{1.5} \cdot N_{\rm i}^{-1} \right)$  \alb

  NO$^+$         & $ \frac{1}{2}{\rm avgh} \left(N_{\rm n}^{-1} \cdot 1.62 \cdot 10^{23}\cdot T^{-0.5},~~14.3 \cdot m_{\rm NO^+}^{-0.5} \cdot T^{1.5} \cdot N_{\rm i}^{-1} \right)$  \alb

    N$^+$        & $ \frac{1}{2}{\rm avgh} \left(N_{\rm n}^{-1} \cdot 1.44 \cdot 10^{23}\cdot T^{-0.5},~~14.3 \cdot m_{\rm N^+}^{-0.5} \cdot T^{1.5} \cdot N_{\rm i}^{-1} \right)$  \alb
         O$^+$        & $ \frac{1}{2}{\rm avgh} \left(N_{\rm n}^{-1} \cdot 1.35 \cdot 10^{23}\cdot T^{-0.5},~~14.3 \cdot m_{\rm O^+}^{-0.5} \cdot T^{1.5} \cdot N_{\rm i}^{-1} \right)$  \alb
    \bottomrule
    \end{tabular*}
    \begin{tablenotes}
      \item[a] Notation and units: $T$ is in Kelvin; $m_{\rm i}$ is the mass of one ion particle in kg; $N_{\rm n}$ is the sum of the neutral species number densities  in 1/m$^3$ and $N_{\rm i}$ is the sum of the positive ion number densities in 1/m$^3$.
      \item[b] The function avgh($a$,$b$) returns the harmonic average of the two arguments as follows $2/(a^{-1}+b^{-1})$.
    \end{tablenotes}
   \end{threeparttable}
\end{table}

\subsection{{High Electric Field Correction\tnote{a}}}

{When ion velocities significantly exceed thermal velocities, the ion drift velocity scales as $(E/N)^{0.5}$, with charge-exchange or hard sphere-type cross sections playing a dominant role [\cite{misc:1968:sinnott, pr:1954:wannier}]. This behavior occurs when the electric field is strong enough that collisions are primarily driven by drift motion rather than thermal motion. We incorporate this effect into our modeling through a curve fit to experimental drift velocity data for positive ions from \cite{misc:1968:sinnott}. Such ``high electric field corrections'' are outlined in Table~\ref{tab:correctedmobilitiesEstar} and are applied to both the GY and SE ion mobilities.}

\begin{table}[!t]
  \center\fontsizetable
  \begin{threeparttable}
    \tablecaption{{Electric field correction for ion mobilities (GY and SE models)\tnote{a}.}}
    \label{tab:correctedmobilitiesEstar}
    \fontsizetable
    \begin{tabular*}{\columnwidth}{c@{\extracolsep{\fill}}c}
    \toprule
    Ion & Corrected mobility, $\rm m^2\cdot V^{-1}\cdot s^{-1}$ \\
    \midrule
  N$_2^+$         & $\min\left( \mu_{\rm N_2^+}, N^{-1} \cdot 2.03\cdot 10^{12}\cdot (E^\star)^{-0.5}\right)$  \alb

      O$_2^+$         & $\min\left( \mu_{\rm O_2^+}, N^{-1} \cdot 3.61\cdot 10^{12}\cdot (E^\star)^{-0.5}\right)$  \alb
      NO$^+$         & $\min\left( \mu_{\rm NO^+}, N^{-1} \cdot 4.47\cdot 10^{12}\cdot (E^\star)^{-0.5}\right)$  \alb

      Other positive ions         & $\min\left( \mu_{\rm i}, N^{-1} \cdot 0.55\cdot (m_{\rm i}E^\star)^{-0.5}\right)$  \alb

    \bottomrule
    \end{tabular*}
        \begin{tablenotes}
      \item[a] Notation and units: $\mu_k$ is the uncorrected ion mobility of species $k$ with units of $\rm m^2\cdot V^{-1}\cdot s^{-1}$; $m_{\rm i}$ is the ion mass in kg; $E^{\star}$ is the reduced electric field ($E^\star \equiv |\vec{E}|/N$) with units of $\rm V\cdot m^2$ and $N$ is the number density of the mixture in $\rm 1/m^3$. 
    \end{tablenotes}
   \end{threeparttable}
\end{table}

\section{Numerical Methods}

The physical model outlined above is implemented in our in-house code CFDWARP (Computational Fluid Dynamics, Waves, Reactions, Plasmas), using the stiffness-free recast of the electric field potential equation of \cite{book:2022:parent}. The convective derivatives are discretized using the  \cite{jcp:1981:roe} scheme turned second-order by means of the Monotone Upwind Scheme for Conservation Laws (MUSCL) approach and the Van Leer Total Variation Diminishing (TVD) limiter. A positivity preserving filter derived in \cite{aiaaconf:2019:parent} is deployed here to avoid negative densities while the eigenvalue conditioning scheme based on the Peclet number recommended by \cite{aiaa:2017:parent} is used to avoid carbuncles. Convergence to steady-state is obtained through a diagonally-dominant alternate direction block-implicit method (DDADI) by \cite{aiaaconf:1987:bardina}, while the potential equation is solved using a combination of the iterative modified approximate factorization (IMAF) scheme by \cite{cf:2001:maccormack} and successive over relaxation (SOR) from \cite{misc:1955:douglas}.

\begin{figure}[ht]
     \centering
     \subfigure[Total thermal conductivity]{\includegraphics[width=0.35\textwidth]{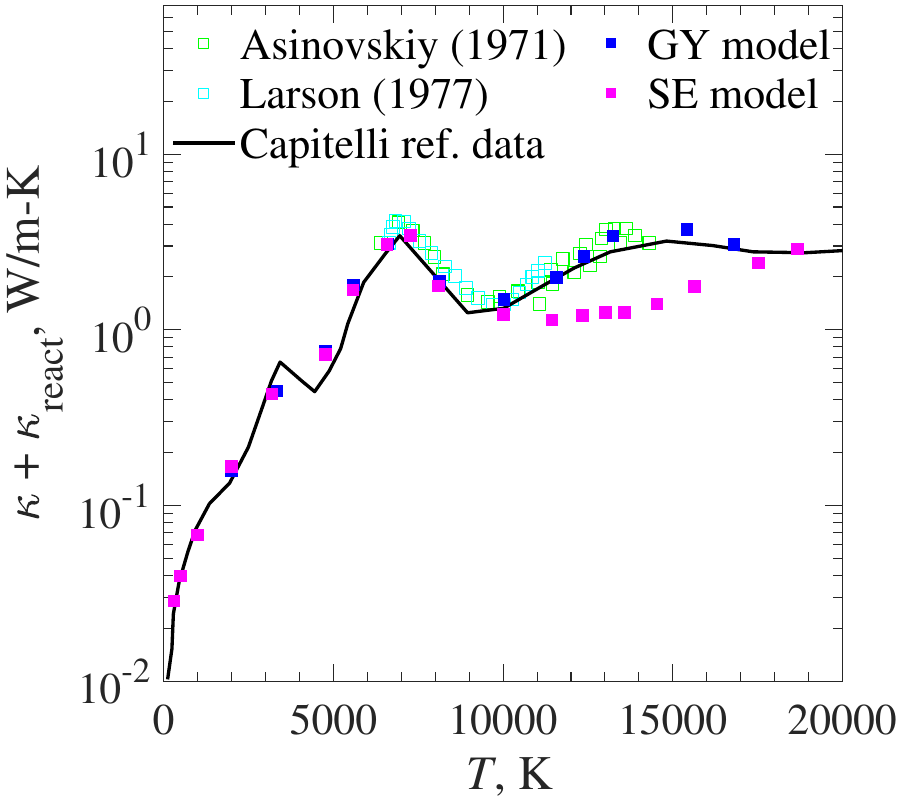}}
     \subfigure[Reactive thermal conductivity]{\includegraphics[width=0.35\textwidth]{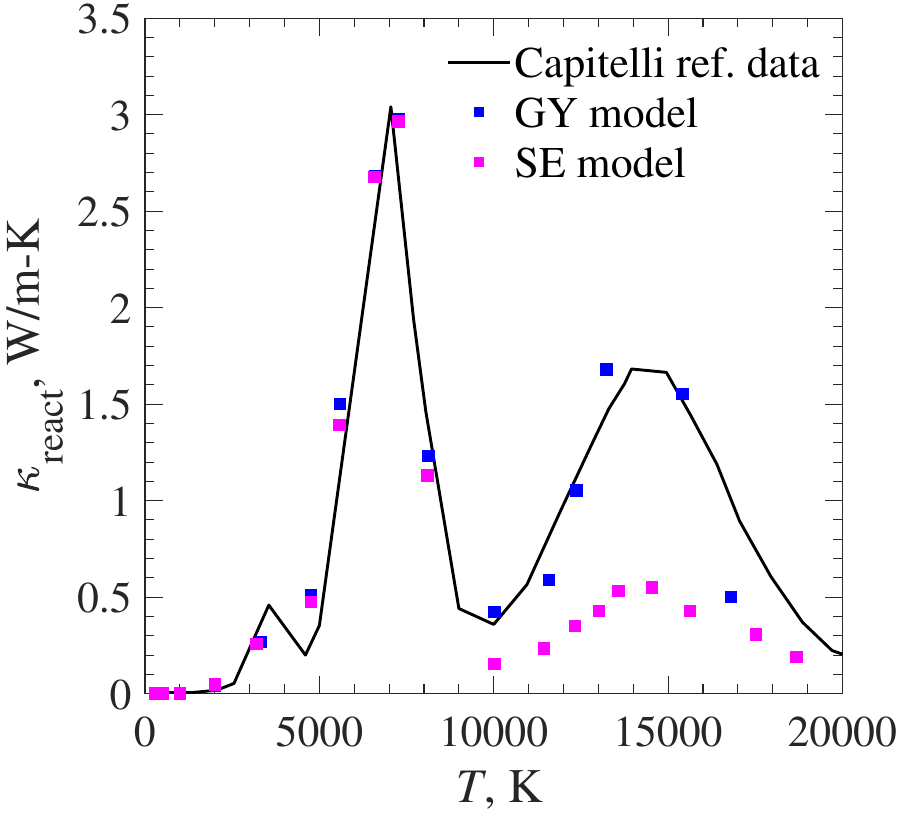}}

\figurecaption{{Comparison of ion mobility models with experimental data of (a) total thermal conductivity and (b) reactive thermal conductivity of equilibrium air at 1 atm.}}
     \label{fig:validation_kappareact_total}
\end{figure}

\begin{figure*}[!ht]
\centering
     \subfigure[$\rm N_2^+$]{\includegraphics[width=0.28\textwidth]{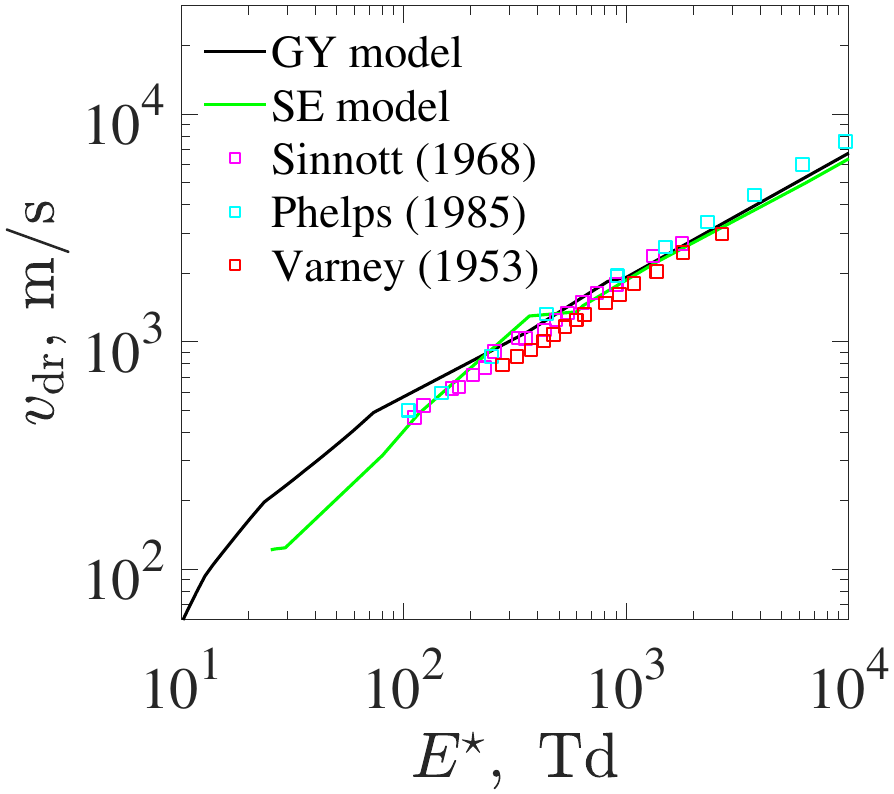}}
     ~~~~~~~\subfigure[$\rm O_2^+$]{\includegraphics[width=0.28\textwidth]{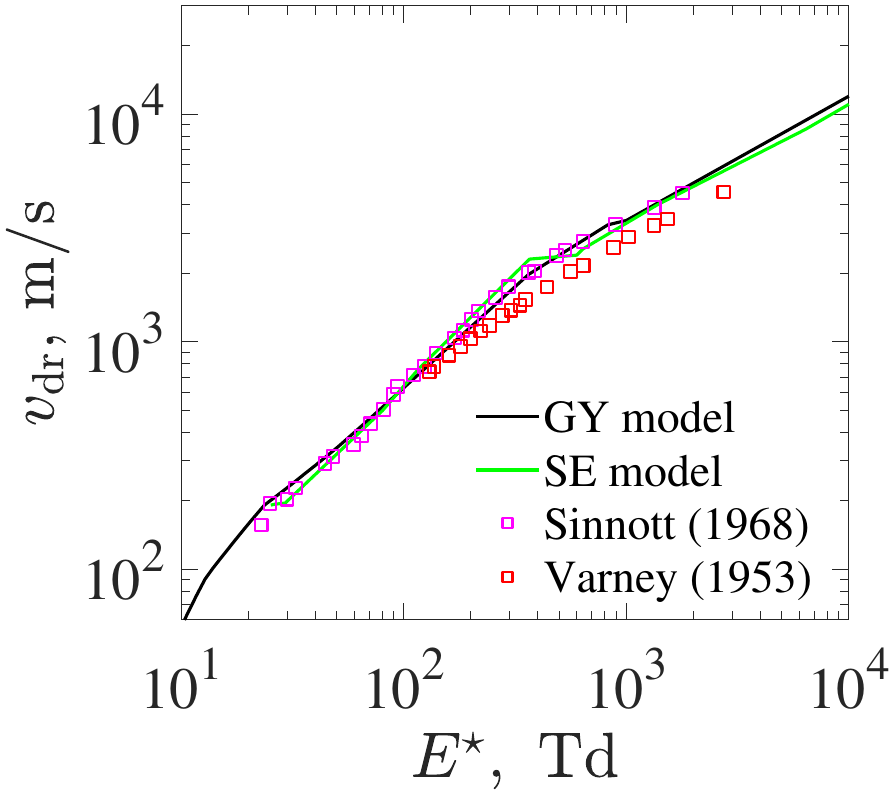}}
     ~~~~~~~\subfigure[$\rm NO^+$]{\includegraphics[width=0.28\textwidth]{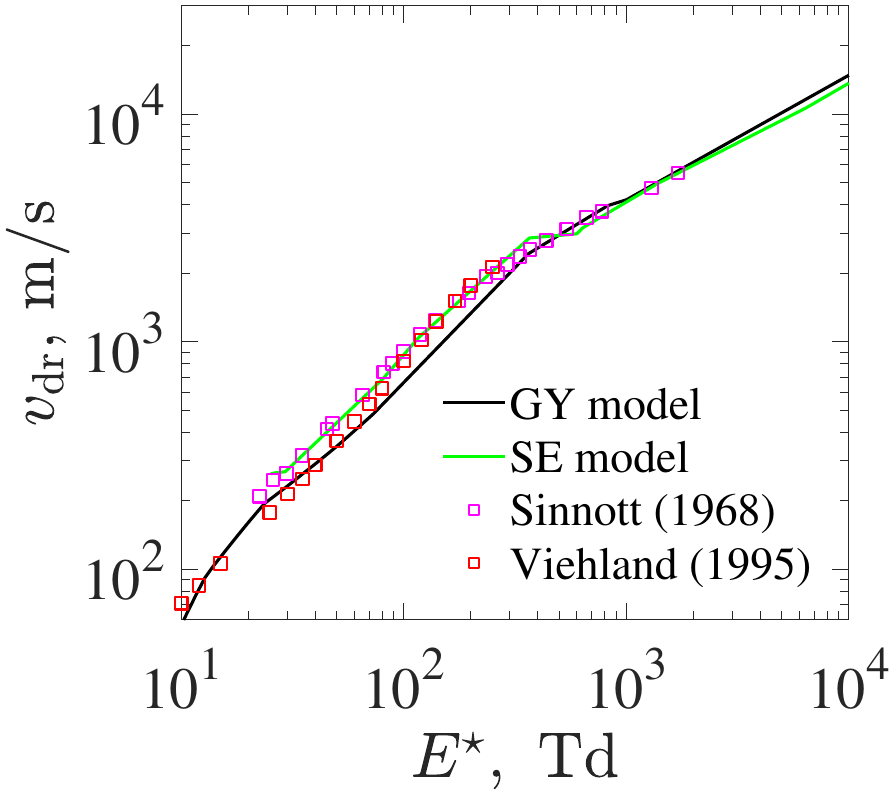}}
      \figurecaption{{Comparison of ion mobility models with experimental data of drift velocity in air at 300 K for (a) $\rm N_2^+$ ions, (b) $\rm O_2^+$ ions, and (c) $\rm NO^+$ ions.}}
     \label{fig:drift_velocity_GY_SE}
\end{figure*}

\section{Validation of Ion Mobility Models}

We present here two validation cases to test the accuracy of both ion mobility models on the basis of (i) reactive thermal conductivity in high temperature air and (ii) ion drift velocity  in low temperature air. 

\subsection{Reactive Thermal Conductivity in High Temperature Air}

{
One way to validate the ion mobility models is through the \emph{reactive thermal conductivity} of an air plasma in equilibrium. This can be explained as follows. First note that the total heat flux in a chemically reaction mixture consists of the sum of the thermal heat flux and the mass diffusion heat flux, as follows:
\begin{equation}
    q_i = -\kappa\dfrac{\partial T}{\partial x_i} - \sum_{k=1}^{n_{\rm s}} \nu_k (h_k+h_{k}^\circ) \frac{\partial w_k}{\partial x_i}
     \label{eqn:heatflux}
\end{equation}
We can rewrite the latter as:
\begin{equation}
    q_i = -(\kappa+\kappa_{\rm react})\dfrac{\partial T}{\partial x_i} 
     \label{eqn:heatflux}
\end{equation}
with the reactive thermal conductivity  given by:
\begin{equation}
    \kappa_{\rm react}=\sum_{k=1}^{n_{\rm s}} \nu_k (h_k+h_{k}^\circ) \frac{\partial w_k}{\partial T}
     \label{eqn:kappa_reactive}
\end{equation}
The reactive thermal conductivity of equilibrium air as a function of temperature is determined as follows. Time-accurate simulations are conducted using the air chemical solver outlined in \cite{jtht:2023:parent} in a container with adiabatic walls until a steady state is reached. The initial gas composition is set to sea-level air, and the initial pressure is adjusted through trial and error to achieve a steady-state pressure of 0.95–1.05 atm. By varying the initial temperature, equilibrium air properties are obtained over the temperature range of 1000–20,000 K at approximately 1 atm. The resulting equilibrium composition is then used to calculate reactive thermal conductivity using Eq. (\ref{eqn:kappa_reactive}) with the GY and SE ion mobility models. The results, shown in Fig.~\ref{fig:validation_kappareact_total}, are compared with experimental data from \cite{tvt:1969:asinovskiy} and \cite{git:1977:larson} and a theoretical curve fit from \cite[page 257, Fig. 10.8]{book:2013:capitelli}.
}

{Interestingly, both total and reactive conductivity exhibit a strong dependence on the ion mobility model at air temperatures above 10,000~K. This is due to ion and electron diffusion being governed by the ambipolar diffusion coefficient, which is directly proportional to ion mobility. As air becomes highly ionized at elevated temperatures, electron and ion  diffusion dominates the reactive heat flux, making it highly sensitive to the ion mobility model used in the calculations.}
 
{The reactive and total thermal conductivities calculated using the GY model show excellent agreement with experimental data across the entire temperature range considered. However, the reactive conductivity predicted using the SE ion mobilities is two to three times smaller than experimental values at temperatures above 10,000~K. This indicates that ion mobility models derived from cross-sectional data perform better at high temperatures than those based on swarm experiments conducted in room-temperature air.
}

\subsection{Ion Drift Velocity in Low Temperature Air}

{Ion mobilities can also be validated by comparing them to drift velocity data in air at sea-level temperature. Experimental drift velocity data for N$_2^+$, O$_2^+$, and NO$^+$ ions from \cite{pr:1953:varney}, \cite{pr:1985:phelps}, and \cite{adnd:1995:viehland}, are compared here with our ion mobility models under the same gas conditions. }

{As shown in Fig.~\ref{fig:drift_velocity_GY_SE}, the SE model agrees well with experimental data across the entire range of reduced electric fields for all ions. The GY model also performs well for reduced electric fields above 400~Td, where the high-field correction to ion mobilities takes effect, resulting in similar predictions for both models. However, at lower reduced electric fields, the GY model shows significant discrepancies with experimental data, particularly for N$_2^+$ (70\% error).}

\section{Hypersonic Waverider Results}

\begin{figure}[ht]
     \centering
     \includegraphics[width=1.0\columnwidth]{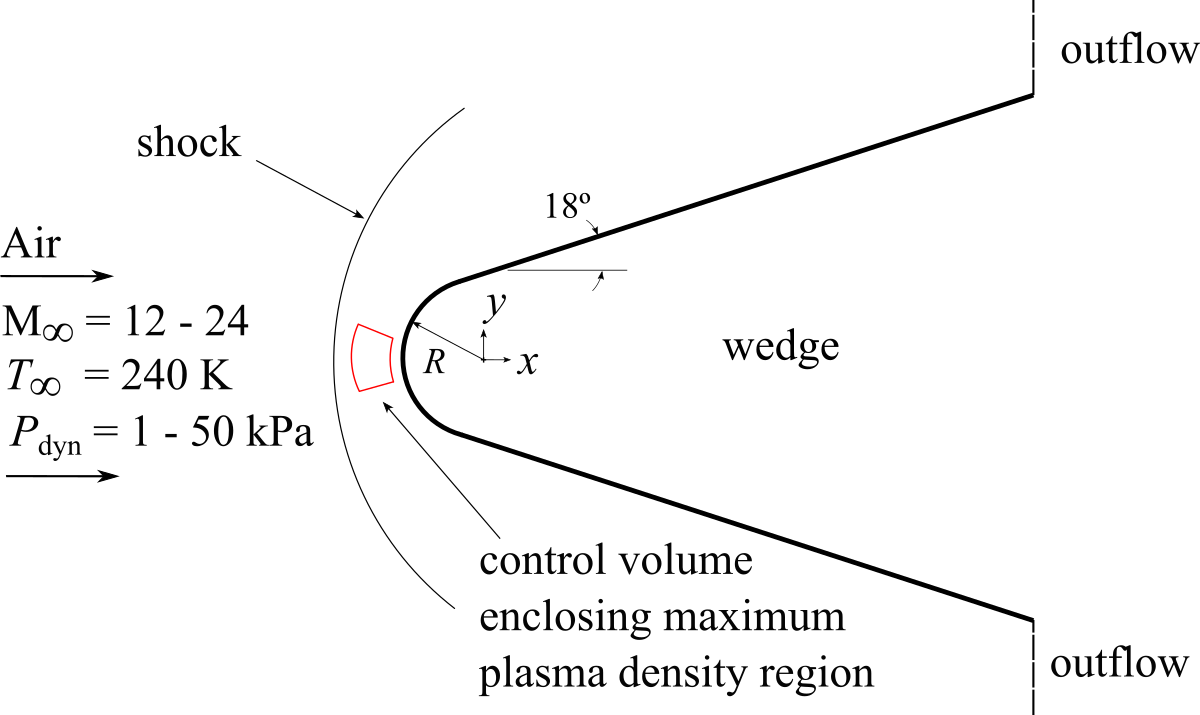}
     \figurecaption{Problem setup for the hypersonic waverider case.}
     \label{fig:schematic}
\end{figure}

The first problem chosen to study the effect of ion mobility on plasma density in detail consists of the hypersonic flow around a waverider vehicle with a sharp leading edge as shown in Fig.~\ref{fig:schematic}. The flow turning angle is 18$^\circ$ and the leading edge has a radius $R$ that can be varied from 1 mm to 20 mm. Such geometry is typical for hypersonic waveriders aiming to reduce drag and aerodynamic heating. The flight Mach number is varied from 12 to 24, and the static inflow temperature is set to 240 K. Non-slip wall boundary conditions are imposed on the waverider surfaces with a wall temperature set to 1400 K. To assess how much the ion mobility model can affect plasma properties at various altitudes, we vary the flight dynamic pressure  within the 1-50 kPa range. Because the dynamic pressure depends on both freestream speed and density (i.e.\ $P_{\rm dyn} = \frac{1}{2}\rho_\infty q_{\infty}^2$ with $q_{\infty}$ the freestream flow speed), a change in dynamic pressure for a given Mach number thus leads to a change in altitude.

\begin{figure*}[ht]
     \centering
     \subfigure[]{\includegraphics[width=0.31\textwidth]{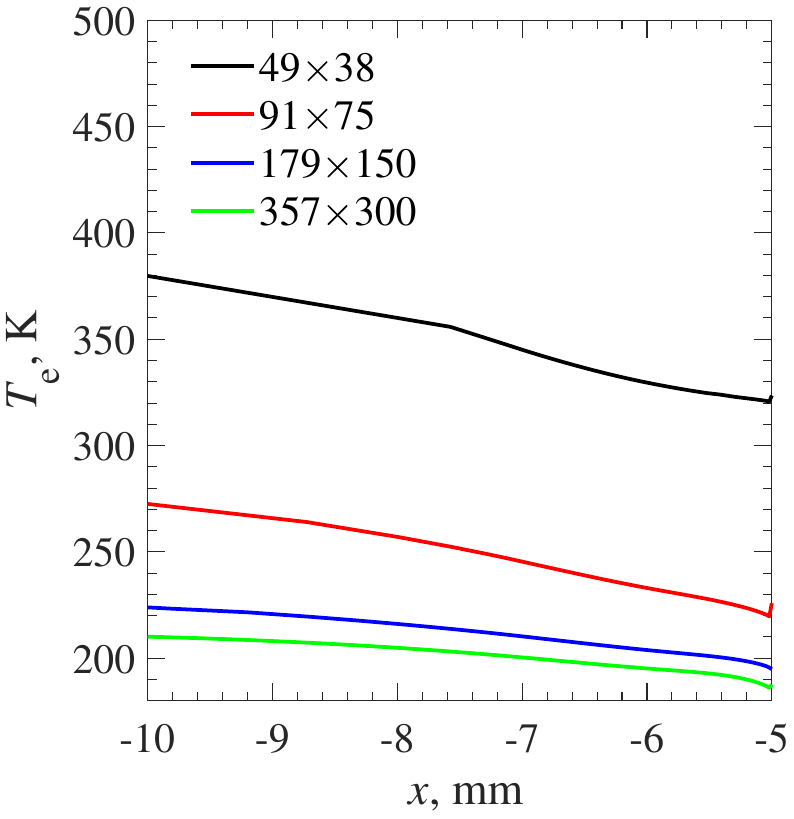}}
     ~~~~~~~~~\subfigure[]{\includegraphics[width=0.31\textwidth]{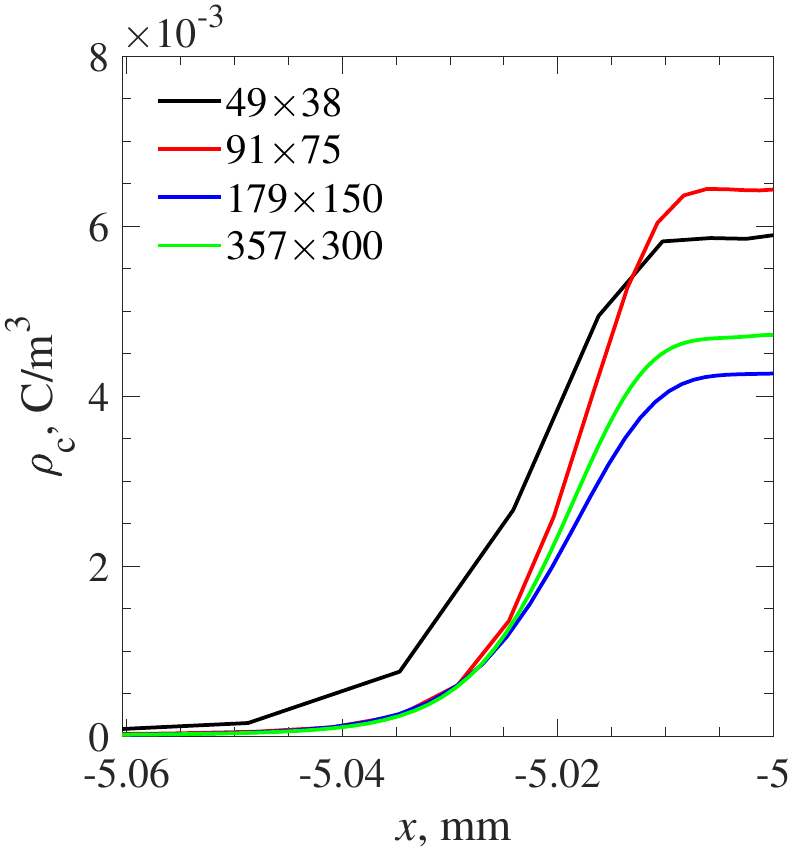}}
     \figurecaption{Grid convergence study for the waverider case using the GY model on the basis of (a) electron temperature along the stagnation line and (b) charge density within the plasma sheath.}
     \label{fig:te_rhoc_griconvergence}
\end{figure*}

An assessment of grid-induced error on the flow properties is presented here for the case  $P_{\rm dyn} =1$ kPa, M = 24 and $R=5$ mm and using four grid levels of $49\times38$, $91\times75$, $179\times150$ and $357\times300$ nodes. Of all measured properties along the stagnation line, electron temperature and charge density exhibit the most sensitivity to the grid. Figure\ \ref{fig:te_rhoc_griconvergence} shows the effect of grid resolution on the electron temperature and charge density profile extracted along the stagnation line for the GY model. {We employ here the Grid Convergence Index (GCI) outlined in \cite{aiaa:1998:roache}. From the results obtained, the observed order of accuracy on the basis of electron temperature is 1.6 and is consistent across the 4 meshes thus indicating that the solution is within the asymptotic range of convergence. Using the GCI, the error associated with the finest $357\times300$ mesh is estimated to be 4\%.  Although not shown here, similar trends are also found when using the SE ion mobility model and for other properties such as the net charge density.}

\begin{figure*}[!ht]
\centering
     \subfigure[$P_{\rm dyn}=1~\rm kPa$]{\includegraphics[width=0.24\textwidth]{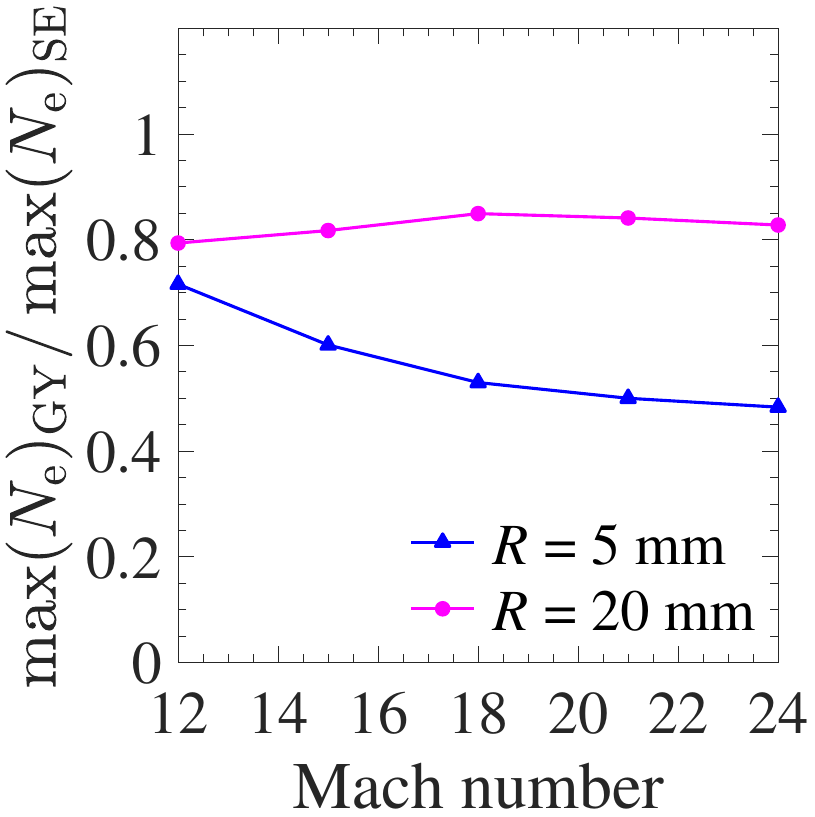}}
     ~~~~~~~~\subfigure[$P_{\rm dyn}=10~\rm kPa$]{\includegraphics[width=0.24\textwidth]{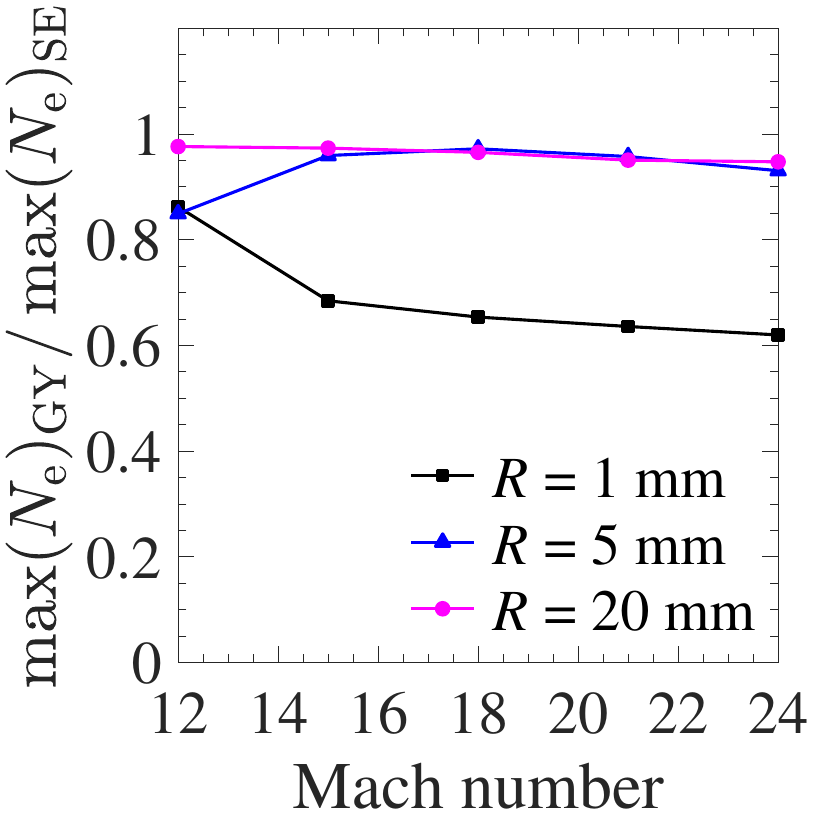}}
     ~~~~~~~~\subfigure[$P_{\rm dyn}=50~\rm kPa$]{\includegraphics[width=0.24\textwidth]{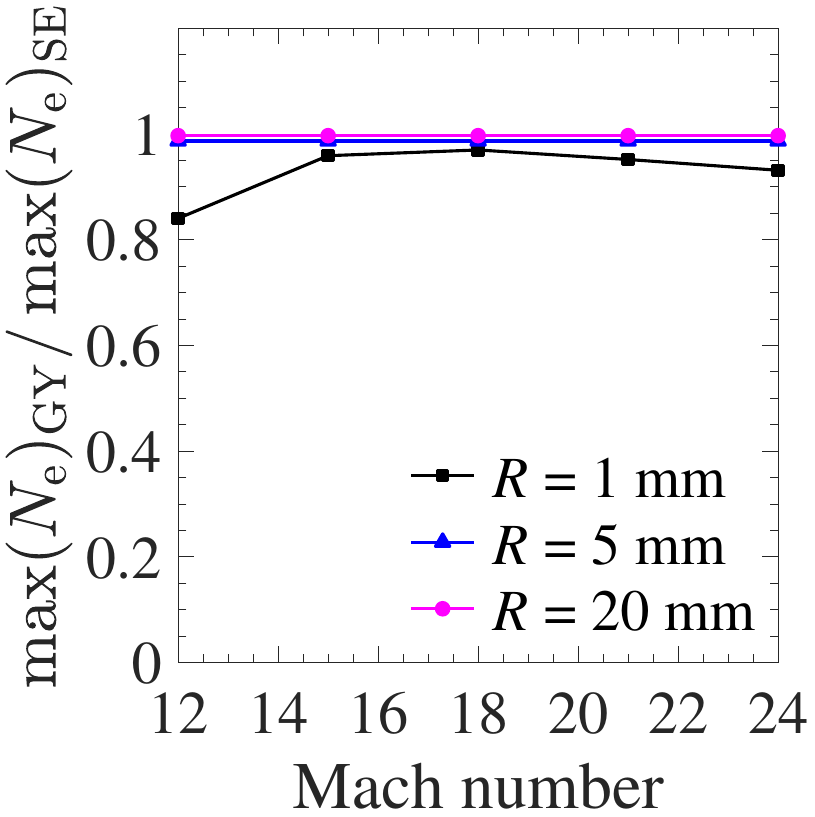}}
      \figurecaption{Effect of flight dynamic pressure $P_{\rm dyn}$, freestream Mach number, and leading edge radius $R$ on the ratio of peak GY electron density and peak SE electron density.}
     \label{fig:parametric_Pdyn_R_Mach}
\end{figure*}

The first step in assessing the sensibility of plasma density to ion mobility involves identifying under which flight conditions the differences in ion mobility models lead to a large difference in the density of the plasma. To do this, we will vary i) the freestream dynamic pressure, ii) the radius of the leading edge, effectively scaling the waverider and iii) the freestream Mach number. Solutions are then obtained for combinations of these parameters. Next, we compute the ratio of peak electron density found using the mobility expressions as given in the GY model, and that obtained using the SE model. The results are shown in Fig.~\ref{fig:parametric_Pdyn_R_Mach} where it can be seen that the largest impact of the ion mobility model on the plasma density occurs when the leading edge radius is low, the flight dynamic pressure is low, and the flight Mach number is high. We will now focus on the test case where the impact of the ion mobility model is the highest (i.e., for which $P_{\rm dyn}=1~$kPa, $M_\infty=24$ and $R=5$~mm) and explain in detail the various physical mechanisms through which the ion mobility does impact plasma density.

\subsection{Effect of Ion Mobility on Electron Loss by Ambipolar Diffusion}

 The first physical mechanism through which ion mobility could impact plasma density is by electron loss to the surface through ambipolar diffusion. Indeed, because the electron density at the surface is several orders of magnitude less than the one in the nearby plasma, and because most of the plasma is quasi-neutral, the electrons diffuse towards the surface according to the ambipolar diffusion coefficient which scales with the ion mobility. To assess the importance of electron loss through ambipolar diffusion we focus on a control volume located in the post-shock stagnation region (see Fig.~\ref{fig:schematic}). This control volume is chosen so that it encloses the region of peak electron density. The various electron gains and losses associated with this control volume are tabulated in Table~\ref{tab:gainslosses_Pdyn_integrals}, listing contributions from chemical reactions occurring within the control volume, convection of electrons through the control volume boundaries and diffusion of electrons (ambipolar diffusion). In the latter, the ambipolar diffusion coefficient depends on the mobility of positive ions, $\mu_{\rm i}$, which is obtained by taking a molar-fraction-weighted average of all positive ion mobilities. As expected, it can be observed that the electron gain from chemical reactions matches the loss of electrons due to the sum of convection and ambipolar diffusion through the volume boundaries, within a small 5\% numerical error.

\begin{table}[t]
\center\fontsizetable
  \begin{threeparttable}
\tablecaption{Electron gains and losses (in kg/s per unit depth) within the specified control volume associated with the waverider case.}
\renewcommand{\arraystretch}{1.4}
\begin{tabular*}{\columnwidth}{@{}l@{\extracolsep{\fill}}cc@{}}
\toprule
Electron gain-loss mechanism  & GY model   & SE model \\
\midrule
$\int_V W_{\rm e} \text{d}V$~(chemical reactions)   & $2.940\cdot10^{-13}$ & $2.839\cdot10^{-13}$\\
$\int_{S} \rho_{\rm e} \vec{V} \cdot\vec{n}\text{d}S$~(convection through boundaries)    & $-5.794\cdot10^{-14}$ &$-1.953\cdot10^{-13}$\\
$\int_{S}  -\frac{\rho \mu_{\rm i} k_{\rm B} T}{|C_{\rm i}|}\left( 1+\frac{T_{\rm e}}{T}\right) \vec{\nabla} w_{\rm e} \cdot\vec{n}\text{d}S$~(ambipolar diff.) & $-2.214\cdot10^{-13}$ & $-1.035\cdot10^{-13}$\\
\bottomrule
\end{tabular*}
\label{tab:gainslosses_Pdyn_integrals}
\end{threeparttable}
\end{table}

\begin{figure*}[!h]
     \centering
     \subfigure[GY model]{\includegraphics[width=0.37\textwidth]{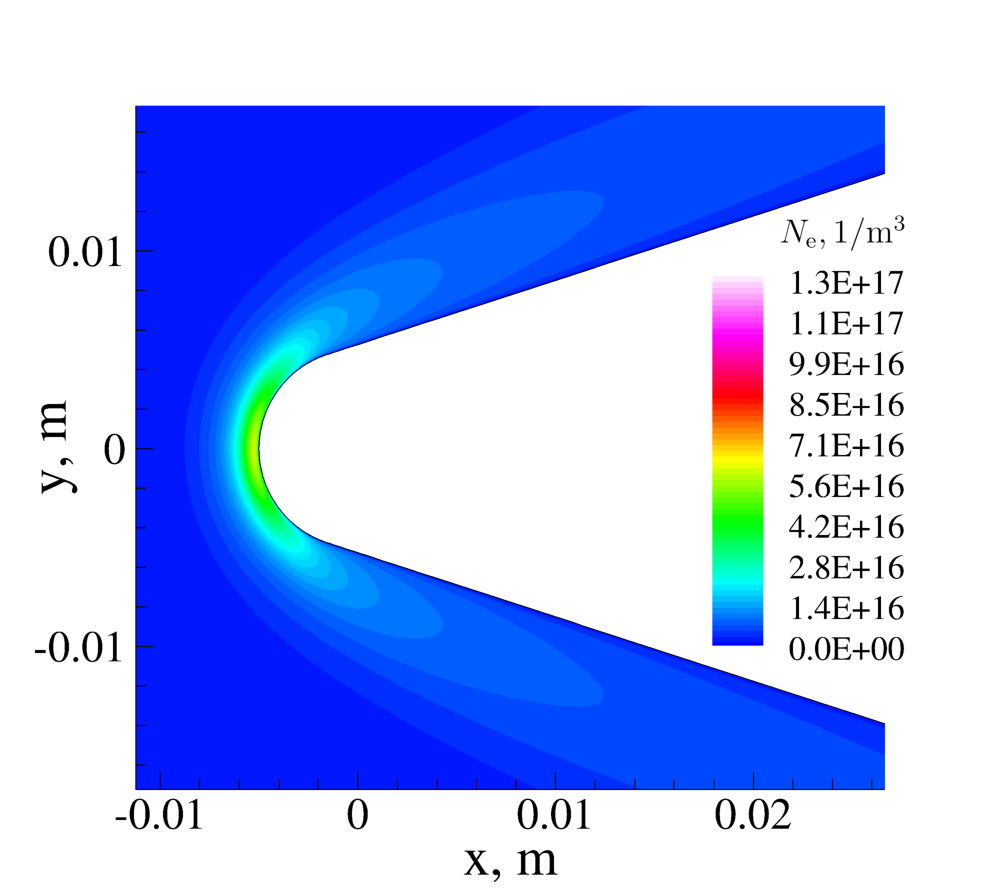}}
     ~~~~~~~\subfigure[SE model]{\includegraphics[width=0.37\textwidth]{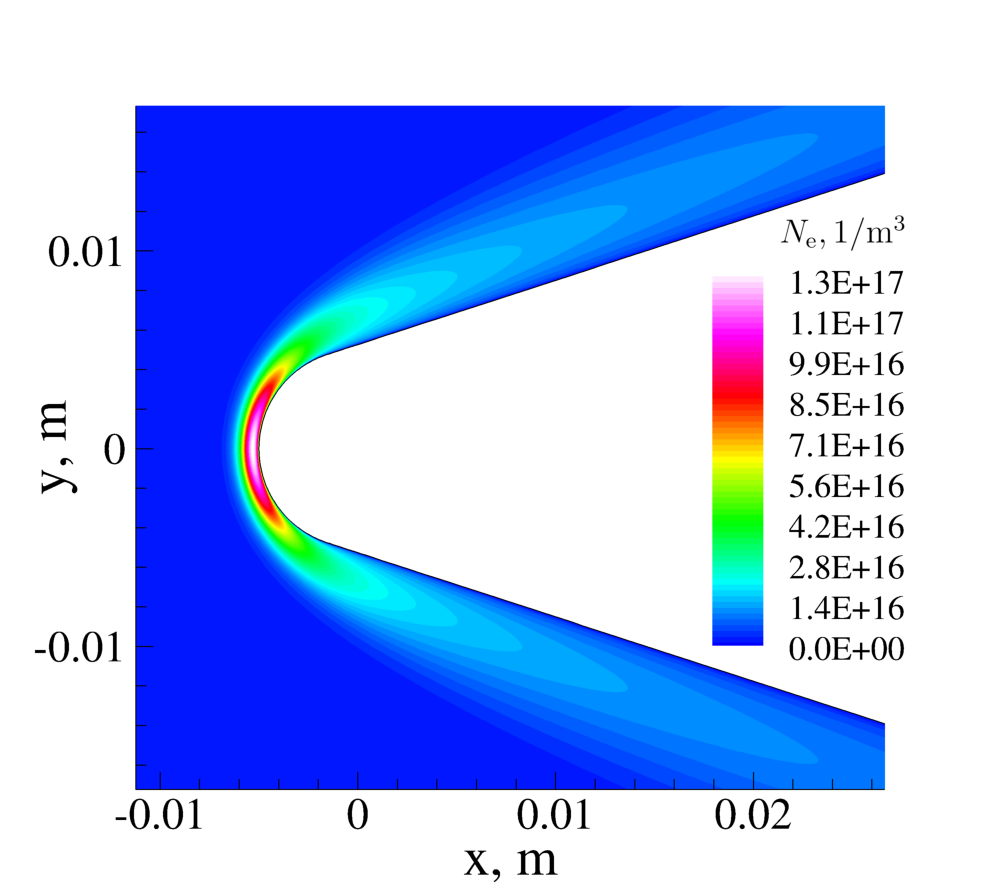}}
     \figurecaption{Comparison of electron number density contours for the waverider case obtained with (a) the GY model and (b) the SE model.}
     \label{fig:Ne_contours_waverider}
\end{figure*}

As shown in Table~\ref{tab:gainslosses_Pdyn_integrals}, electron losses by ambipolar diffusion amount to 36\% of the electron gains when using the SE model, while this percentage increases to 75\% for the solution using the GY model. If the amount of ambipolar diffusion through the boundaries is large relative to the amount of electrons gained from chemical reactions, the electron density peak along the stagnation line should decrease. This is exactly what is observed in Fig.~\ref{fig:Ne_contours_waverider}, where using the GY model decreases the peak of plasma density by a factor of two relative to the SE model solution.

\begin{figure*}[!h]
     \centering
     \subfigure[]{\includegraphics[width=0.38\textwidth]{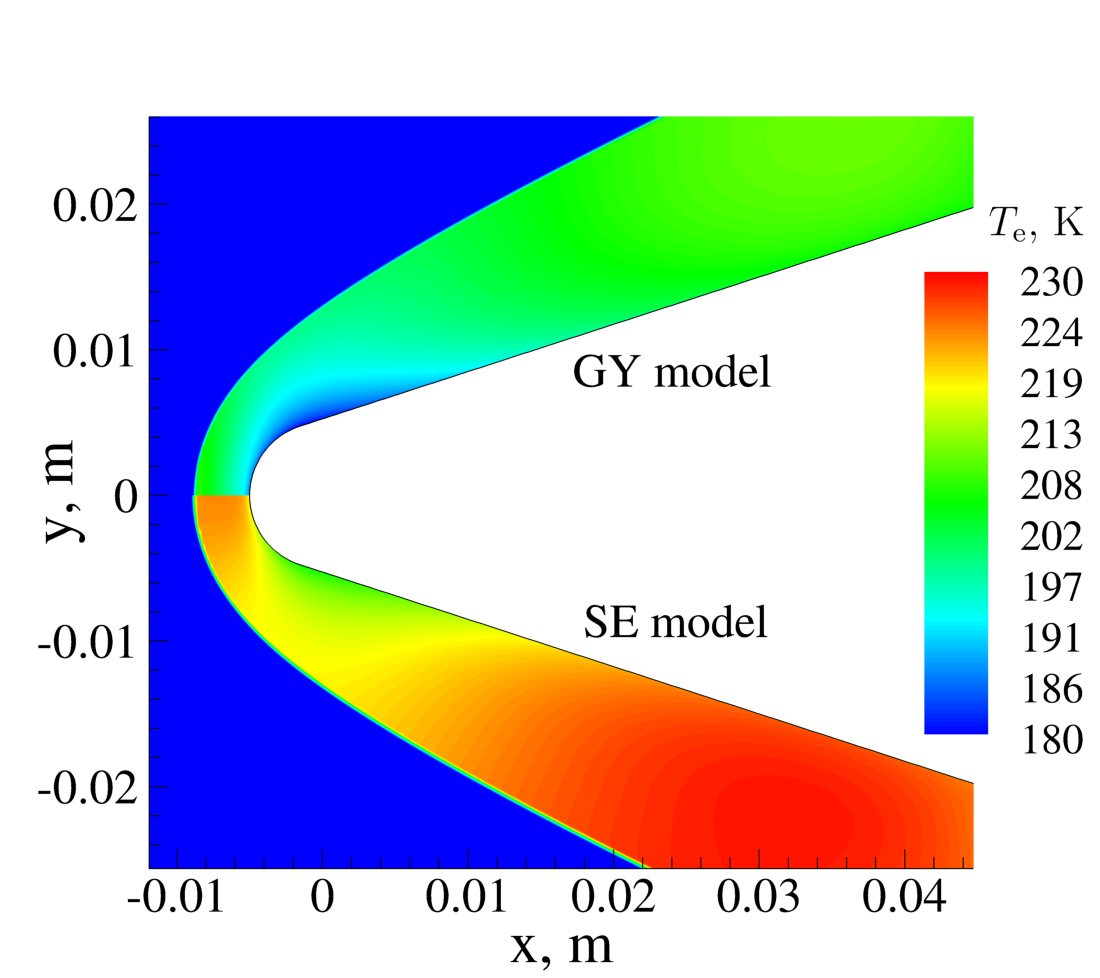}}
     ~~~~~~~\subfigure[]{\includegraphics[width=0.38\textwidth]{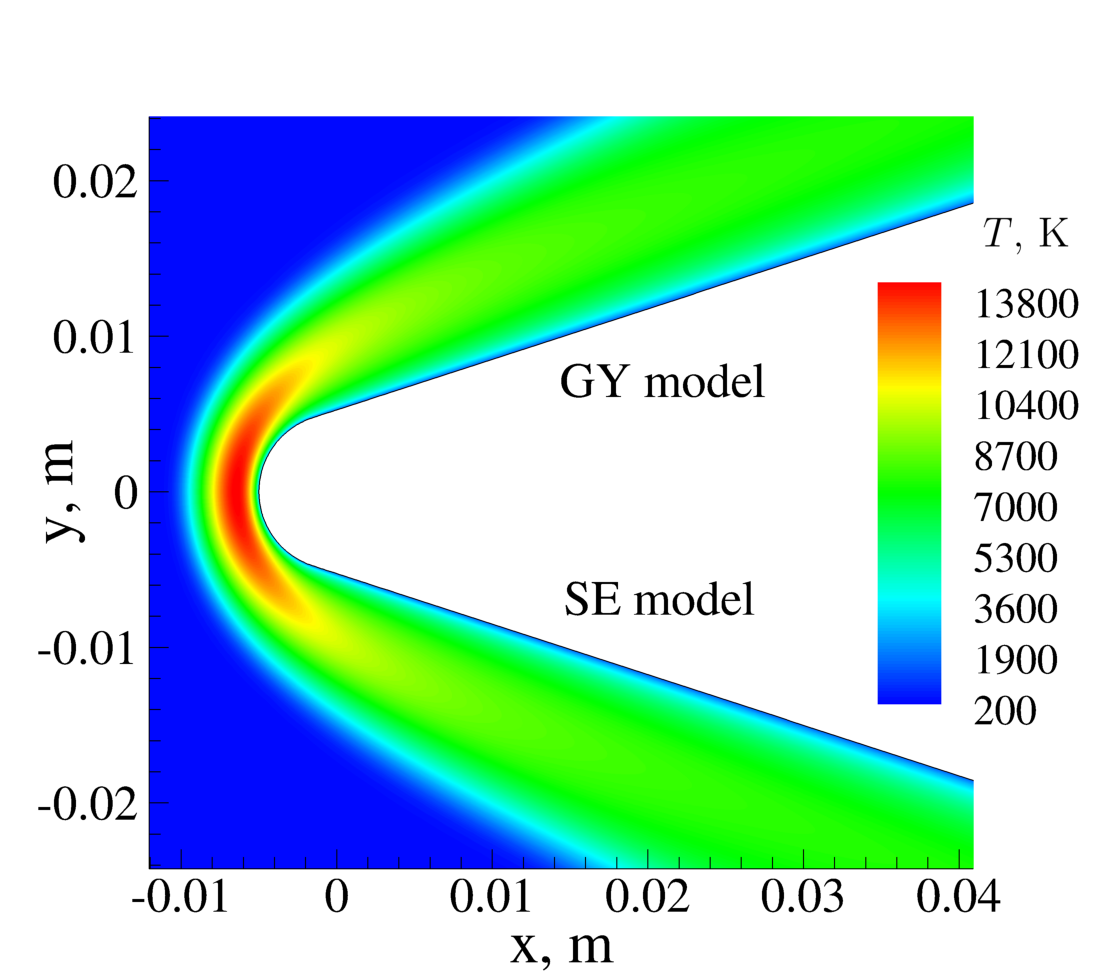}}
     \figurecaption{Effect of ion mobility model on (a) electron temperature {and (b) bulk gas temperature} for the waverider case.}
     \label{fig:Te_T_waverider_1kPa_10kPa_5mm_M24}
\end{figure*}

\begin{figure*}[!ht]
     \centering
     \subfigure[]{\includegraphics[width=0.33\textwidth]{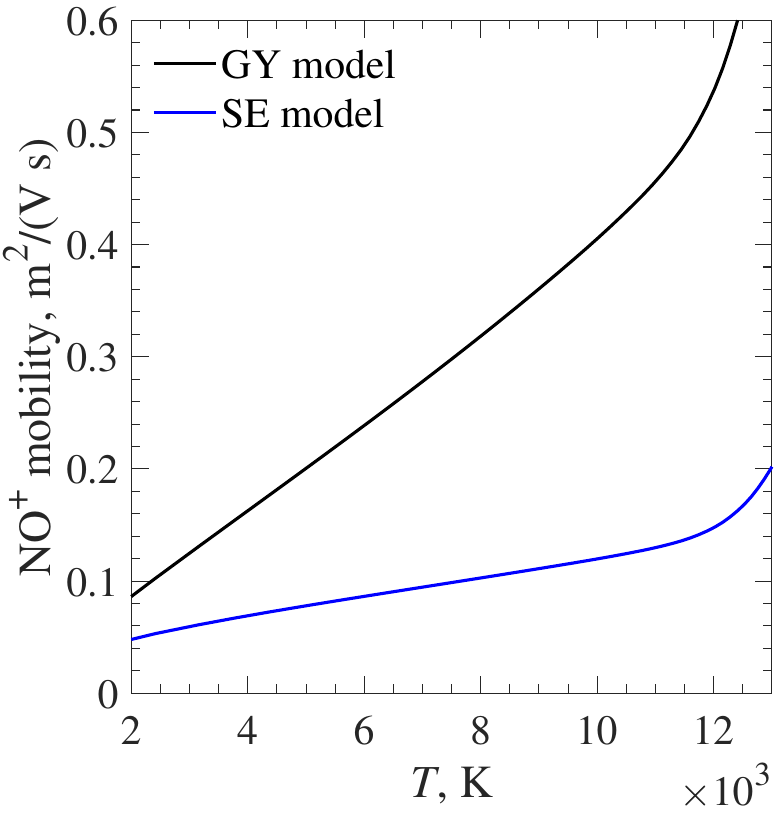}}
     \hspace{1.0cm}
     ~~~~~~~\subfigure[]{\includegraphics[width=0.33\textwidth]{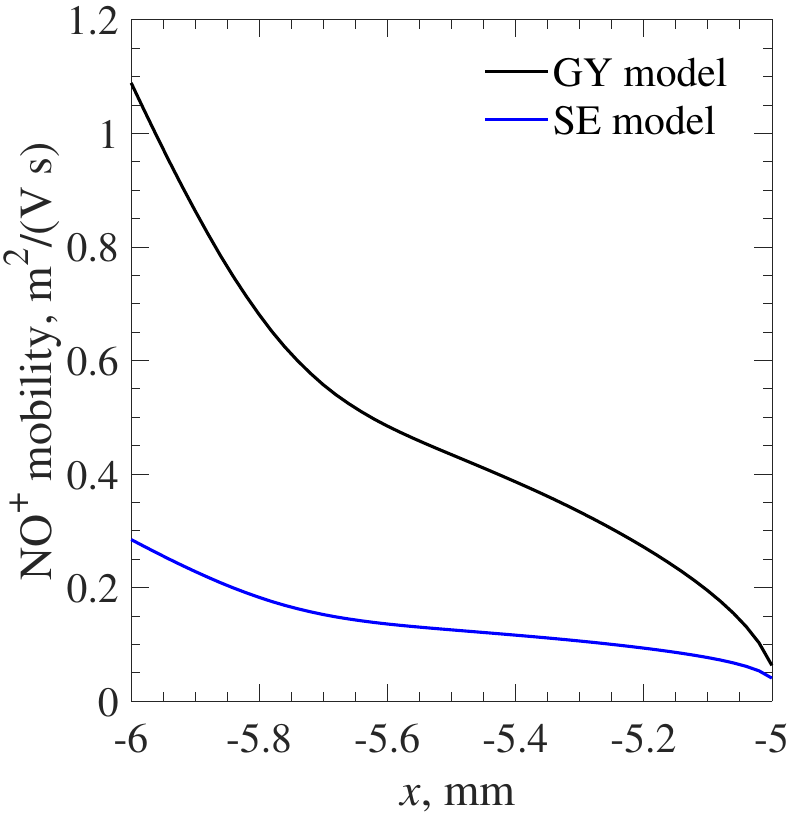}}
     \figurecaption{Effect of ion mobility model on $\rm NO^+$ ion mobility (a) over a range of representative bulk temperature and (b) along the waverider stagnation line.}
     \label{fig:mobility_NOplus_comp}
\end{figure*}

The amount of ambipolar diffusion depends on the ambipolar diffusion coefficient, which is proportional to ion mobility. Electron temperature also has a large impact on ambipolar diffusion because the ambipolar diffusion coefficient scales with $1+T_{\rm e}/T$. However, electron temperature has a small effect on ambipolar diffusion here because the electron temperature remains an order or magnitude less than the gas temperature (see Fig.~\ref{fig:Te_T_waverider_1kPa_10kPa_5mm_M24}). Thus, it is the greater ion mobilities predicted by the GY model that lead to ambipolar diffusion becoming much more relevant relative to electron losses due to recombination reactions within the control volume. Indeed, as shown in  Fig.~\ref{fig:mobility_NOplus_comp}, the dominant ion mobility predicted by the GY model can be 2 to 4 times greater than the one of the SE model.

\subsection{Effect of Ion Mobility on Electron Cooling in Non-neutral Sheaths}

Another way ion mobility can influence plasma density is by affecting the electron temperature. Indeed, as can be seen in Fig.\ \ref{fig:Te_T_waverider_1kPa_10kPa_5mm_M24} the GY model leads to a decrease in electron temperature of 10-20\% compared to the SE model. In turn, a change in electron temperature does affect plasma density by altering the electron-ion recombination rates which depend on electron temperature. The impact of the ion mobility model on electron temperature is peculiar because the ion mobility does not appear explicitly in the electron energy transport equation used to determine electron temperature (see Eq.~(\ref{eqn:ee_transport})).


As will be here demonstrated, this is due to an unexpected impact of ion mobility on the rate of electron cooling through the $\vec{E}\cdot\vec{J}_{\rm e}$ term. Because there is no externally applied electric field, this term will be largest in the plasma sheath because (i) the electric field increases towards the surface as dictated by Gauss' law and (ii) the electron current $J_{\rm e}$ remains high through the sheath due to the large electron density gradient. Further, because the electric field points in the opposite direction of the electron current within the plasma sheath, the $\vec{E}\cdot \vec{J}_{\rm e}$ term leads to \emph{electron cooling} within the sheath. As was first observed by \cite{pf:2021:parent}, the electron cooling in the sheath can be quite considerable and reach values comparable to electron heating due to electron-neutral collisions. Such is the case as well here. Indeed, when comparing electron cooling in the sheath with the electron energy gain and loss integrals over the entire computational domain, we find that over 75\% of the total cooling of electrons takes place in the plasma sheath (see Table \ref{tab:gainslosses_energy}). Rather interestingly, it is seen that the ion mobility model influences the amount of electron cooling in the sheath, with the GY model leading to $\vec{E}\cdot \vec{J}_{\rm e}$ being 11\% smaller than the SE model. This is peculiar because this term does not depend on ion mobility but rather on electron mobility. Clearly, because the electron mobility is the same when using either the SE or GY model, the $\vec{E}\cdot \vec{J}_{\rm e}$ must somehow depend in an unexpected way on ion mobility. 

To find out how the $\vec{E}\cdot\vec{J}_{\rm e}$ term depends on ion mobility, let us start from the flux due to drift and diffusion of electrons and ions given by:
\begin{equation}
    \vec{\Gamma}_{\rm e}= - \mu_{\rm e} N_{\rm e}\vec{E}-\frac{\mu_{\rm e} k_B T_{\rm e}}{|C_{\rm e}|}\vec{\nabla} N_{\rm e} 
    \label{eqn:eflux}
\end{equation}
\begin{equation}
\vec{\Gamma}_{\rm i}= \mu_{\rm i}N_{\rm i}\vec{E}-\frac{\mu_{\rm i} k_B T_{\rm i}}{|C_{\rm i}| }\vec{\nabla} N_{\rm i} 
    \label{eqn:ionflux}
\end{equation}
In a Debye sheath bounded by a dielectric the current perpendicular to the surface is zero and, thus,  the flux of ions is equal to the flux electrons. Setting Eq.~(\ref{eqn:eflux}) equal to Eq.~(\ref{eqn:ionflux}) and solving for the electric field we obtain:
\begin{equation}
    \vec{E} \approx- \dfrac{k_B}{|C_{\rm e}|}\dfrac{ \mu_{\rm e} T_{\rm e} \vec{\nabla} N_{\rm e}  -  \mu_i  T_i \vec{\nabla} N_i  }{\mu_i N_i+  \mu_{\rm e} N_{\rm e}}
    \label{eqn:efield_full}
\end{equation}
Close to the surface, the dielectric surface boundary condition yields:
\begin{equation}
\mu_{\rm e} N_{\rm e} = \gamma_{\rm e}\mu_{\rm i} N_{\rm i}
\label{eqn:boundarysheath}
\end{equation}
with $\gamma_{\rm e}$ the electron secondary emission coefficient. Far from the surface and over most of the sheath region, as the electron and ion densities start to become similar, the electron mobility remains much larger than the ion mobility, so that $\mu_{\rm i} N_{\rm i} \ll \mu_{\rm e} N_{\rm e}$. We thus have two bounds for the denominator of Eq.~(\ref{eqn:efield_full}): $\mu_{\rm e}N_{\rm e}$ over most of the sheath and $\mu_{\rm e}N_{\rm e}/\gamma_{\rm e}$ near the surface where the boundary condition Eq.~(\ref{eqn:boundarysheath}) holds. We can write these bounds in a compact form:
\begin{equation}
\mu_{\rm e} N_{\rm e} + \mu_{\rm i} N_{\rm i} \approx \mu_{\rm e} N_{\rm e} \xi
\label{eqn:bounds1}
\end{equation}
with $\xi$ varying between 1 at the sheath edge and $1/\gamma_{\rm e}$ at the surface.
Now, the $\mu_i  T_i \vec{\nabla} N_i$ term in Eq.~(\ref{eqn:efield_full}) can be neglected in comparison to $\mu_{\rm e} T_{\rm e} \vec{\nabla} N_{\rm e}$, because while the electron temperature can be an order of magnitude lower than the ion temperature, (i) the electron mobility is much larger than the ion mobility and (ii) the ion density gradient is smaller than the electron density gradient because the electron density at the surface is several orders of magnitude less than the ion density, while at the sheath edge these are equal. With these simplifications, we can further approximate the electric field, Eq.~(\ref{eqn:efield_full}), as:
\begin{equation}
\vec{E} \approx -\dfrac{k_B}{|C_{\rm e}|}\dfrac{ T_{\rm e} \vec{\nabla} N_{\rm e}}{\xi N_{\rm e}}
\label{eqn:efield_ambipolar}
\end{equation}
The electron density gradient can be estimated by dividing the change in electron density in the sheath by a length scale $L_{\rm sheath}$ proportional to the sheath thickness. 
Because the electron density is much smaller at the surface and the electron and ion densities at the sheath edge are the same, we can set
\begin{equation}
|\vec{\nabla} N_{\rm e}| \sim \dfrac{N_{\rm e}}{L_{\rm sheath}}
\label{eqn:gradne}
\end{equation}
The length $L_{\rm sheath}$ will scale with the Debye length $\lambda_D$. In a classical sheath where the ion density is larger than the electron density the Debye length scale is:
\begin{equation}
\lambda_D=\sqrt{\frac{\epsilon_0 k_B T}{N_{\rm i} |C_{\rm e}|^2}}
\label{eqn:debyelength}
\end{equation}
Substitution of Eq.~(\ref{eqn:gradne}) into Eq.~(\ref{eqn:efield_ambipolar}) and using Eq.~(\ref{eqn:debyelength}) as $L_{\rm sheath}$ leads to the following order-of-magnitude estimate for the magnitude of the electric field:
\begin{equation}
E \sim \dfrac{T_{\rm e}}{T}\sqrt{\frac{   N_{\rm i} k_B T}{ \epsilon_0\xi^2}}
\label{eqn:Ex}
\end{equation}
\begin{figure*}[!t]
     \centering
     \subfigure[GY model]{\includegraphics[width=0.36\textwidth]{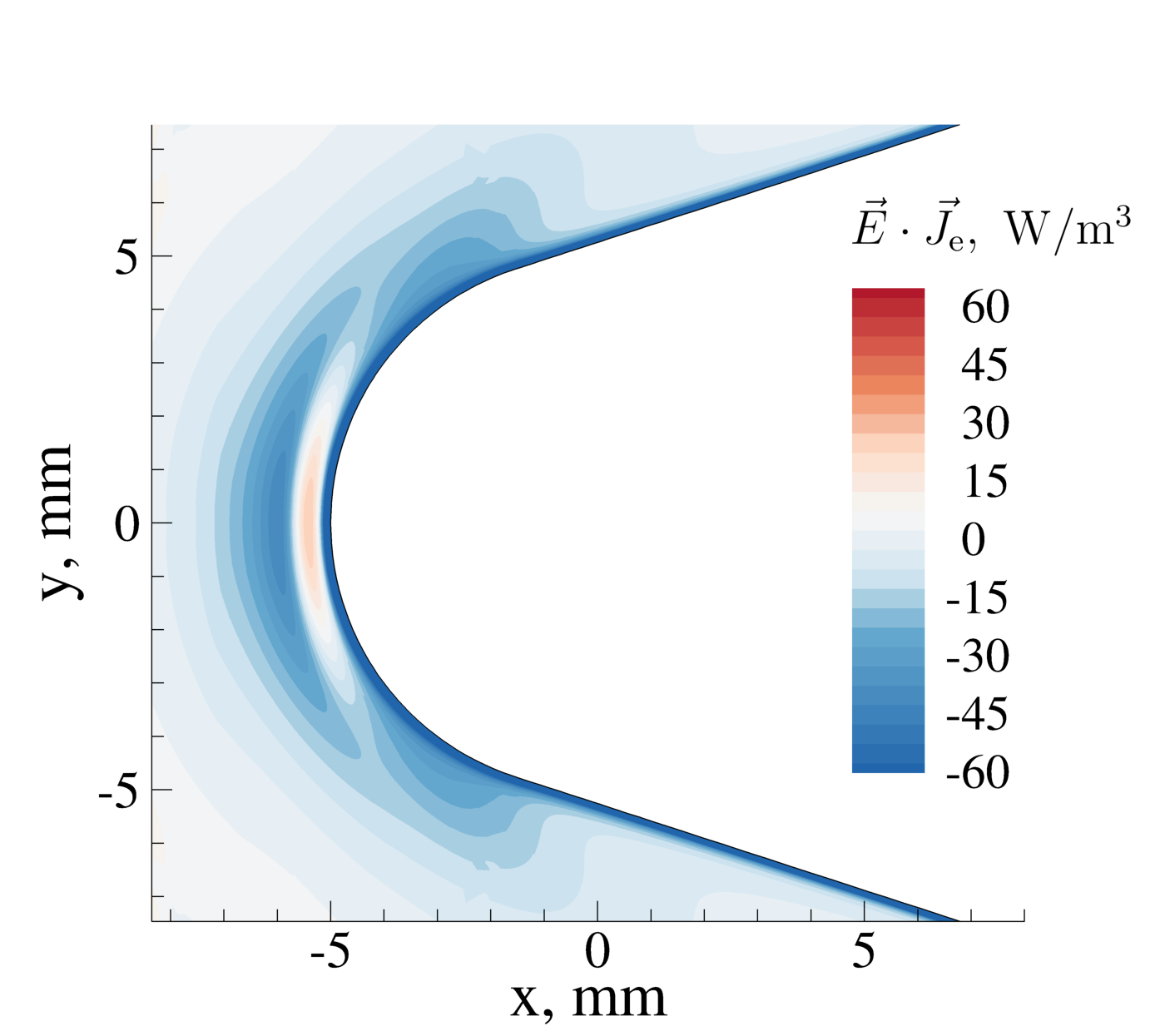}}
     ~~~~~~~~\subfigure[SE model]{\includegraphics[width=0.36\textwidth]{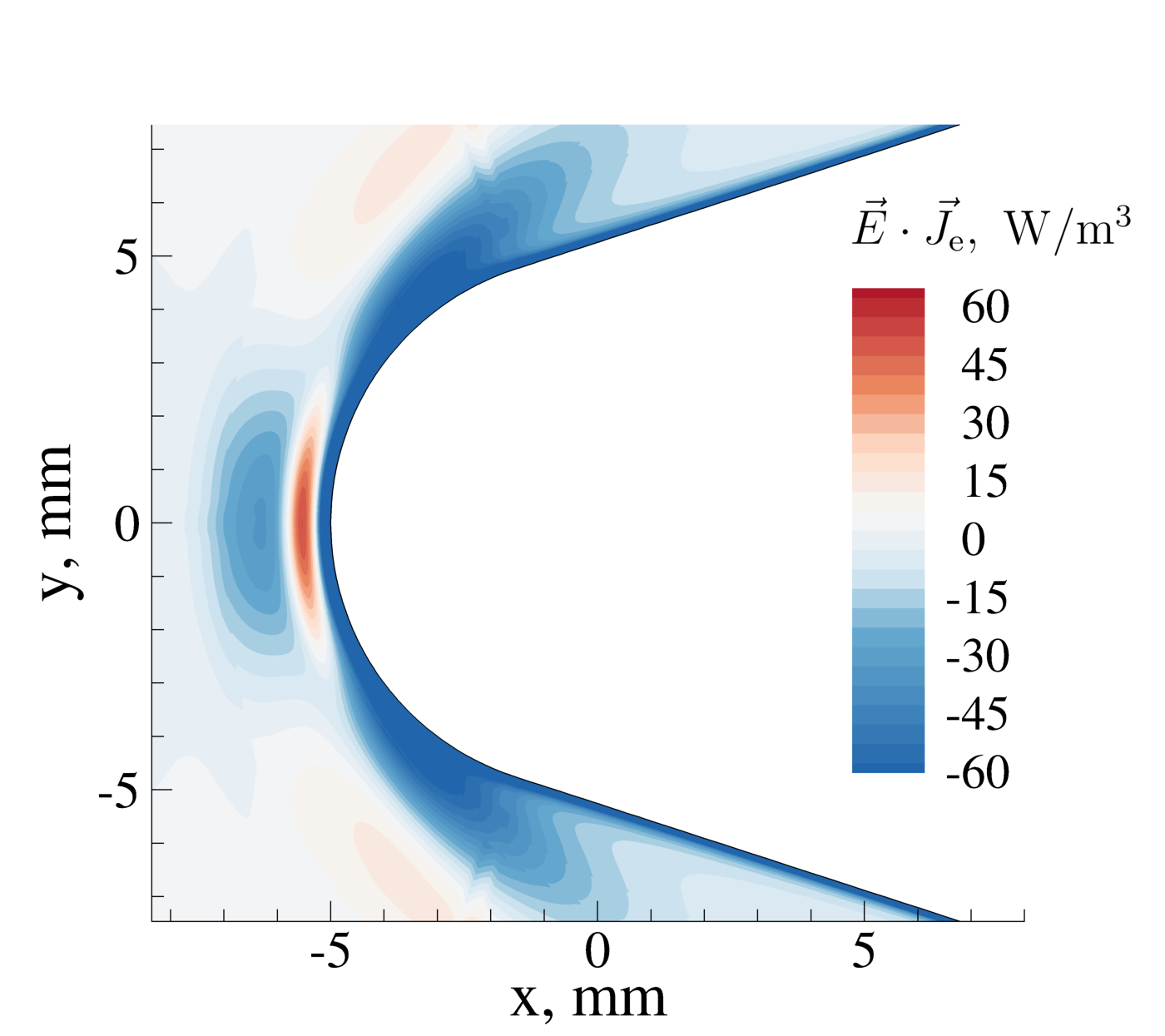}}
     \figurecaption{Electron energy input for the waverider case obtained with (a) the GY model, and (b) the SE model.}
     \label{fig:EdotJe_contours_waverider}
\end{figure*}
To find an estimate for the electron current $J_{\rm e}$ in $\vec{E}\cdot\vec{J}_{\rm e}$, recall that within the Debye sheath the ion and electron current densities are equal, thus
\begin{equation}
{J}_{\rm e} \sim |C_{\rm e}| N_{\rm i}{V}_{\rm i}
\end{equation}
and noting that most of the contribution to the velocity of ions is due to drift, i.e.~${V}_{\rm i} = \mu_i E$, the electron current density magnitude can be estimated as:
\begin{equation}
{J}_{\rm e} \sim |C_{\rm e}|\mu_i N_{\rm i}E
\label{eqn:Je}
\end{equation}
Substituting the electric field estimate Eq.~(\ref{eqn:Ex}) into  Eq.\ (\ref{eqn:Je}) we finally arrive at a scaling law for the electron energy input term $\vec{E} \cdot \vec{J}_{\rm e}$ :
\begin{equation}
\vec{E}\cdot\vec{J}_{\rm e}  \sim  \frac{k_B|C_{\rm e}|  }{\xi^2 \epsilon_0} \mu_i N_{\rm i}^2  T \left(\dfrac{T_{\rm e}}{T}\right)^2 
\end{equation}
We can further integrate the latter through the sheath noting that $T$, $T_{\rm e}$, $\mu_{\rm i}$, and $N_{\rm i}$ do not vary too significantly through the Debye sheath. Then, approximating the sheath length $L_{\rm sheath}$ by the Debye length as outlined in Eq.~(\ref{eqn:debyelength}) we obtain the following scaling law for the integral of the $E \cdot J_{\rm e}$ term within the sheath:
\begin{equation}
\int_{x=0}^{L_{\rm sheath}} (\vec{E}\cdot\vec{J}_{\rm e}) {\rm d} x \sim  
 \frac{k_B^\frac{3}{2} \mu_i N_{\rm i}^\frac{3}{2}   T_{\rm e}^2  }{\xi^2 \sqrt{\epsilon_0 T}}
\label{eqn:edotje_scale}
\end{equation}

\begin{table*}[t]
\center\fontsizetable
  \begin{threeparttable}
\renewcommand{\arraystretch}{1.2}
\tablecaption{Electron energy gain by region in W per unit depth.}
\begin{tabular*}{\textwidth}{@{}l@{\extracolsep{\fill}}cccc@{}}
\toprule
\multirow{2}*{\begin{minipage}{3.5cm}Electron energy gain-loss
surface integral\end{minipage}} & \multicolumn{2}{c}{GY model}& \multicolumn{2}{c}{SE model}\\
\cmidrule(lr){2-3}\cmidrule(lr){4-5}
 & Sheath region\tnote{a}    & Full domain & Sheath region\tnote{a}   & Full domain  \\
\midrule

$\int_{V} (-Q_{\rm e})\text{d}V$    & $0.0007$   &$0.0153$& $0.0008$&$0.0171$\\
$\int_V  \left(\vec{E} \cdot \vec{J}_{\rm e}\right)\text{d}V$   & $-0.0107$    &$-0.0138$&$-0.012$&$-0.01585$\\
$\int_V \left(W_{\rm e}e_{\rm e}\right) \text{d}V$   & $8.8\cdot10^{-7}$    &$0.0035$&$-8.7\cdot10^{-8}$&$0.0038$\\
$\int_V (W_{\rm e} e_{\rm e} + \vec{E}\cdot \vec{J_{\rm e}} - Q_{\rm e})~{\rm d}V$ & $-0.01$  &$0.005$&$-0.011$&$0.0051$\\
\bottomrule
\end{tabular*}
\begin{tablenotes}
\item[{a}]{The plasma sheath is defined as the region adjacent to the surface where the charge density $\rho_c > 10^{-5}~{\rm C/m^3}$.}
\end{tablenotes}
\label{tab:gainslosses_energy}
   \end{threeparttable}

\end{table*}

The expression obtained in Eq.~(\ref{eqn:edotje_scale}) can now be used to explain the discrepancies in electron temperature due to the cooling of electrons when using either ion mobility model. To clearly assess the influence of ion mobility, we divide  $\vec{E}\cdot \vec{J}_{\rm e}$ obtained with the SE model by $\vec{E}\cdot \vec{J}_{\rm e}$ obtained with the GY model. Noting that the gas temperature is not affected much by the ion mobility model, we obtain the following expression:
\begin{equation}
\dfrac{\int_{x=0}^{L_{\rm sheath}} (\vec{E}\cdot\vec{J}_{\rm e})_{\rm SE} {\rm d} x }{\int_{x=0}^{L_{\rm sheath}} (\vec{E}\cdot\vec{J}_{\rm e})_{\rm GY} {\rm d} x}
\sim
\dfrac{\left(N_{\rm i}^\frac{3}{2}  T_{\rm e}^2 \right)_{\rm ~SE}}{\left(N_{\rm i}^\frac{3}{2}  T_{\rm e}^2 \right)_{\rm ~GY}} \cdot \dfrac{\left(\mu_{\rm i} \right)_{\rm ~SE}}{\left(\mu_{\rm i} \right)_{\rm ~GY}}
\label{eqn:edotje_scale_ratio}
\end{equation}

For the case under consideration, the ratio ${\left(N_{\rm i}^{1.5}  T_{\rm e}^2 \right)_{\rm ~SE}}/{\left(N_{\rm i}^{1.5}  T_{\rm e}^2 \right)_{\rm ~GY}}$ can be measured to be approximately 2.2 on average within the sheath region (because the ion density is two times larger and the electron temperature is 10\% higher within the region of the sheath where the electric field is considerable). On the other hand, the second ratio $\left(\mu_{\rm i} \right)_{\rm ~SE}/\left(\mu_{\rm i}\right)_{\rm GY}$ can be measured to be approximately 0.5. Thus, our scaling law predicts that the electron cooling in the sheath is about 10\% higher when using the SE model than when using the GY model. Such is in very close agreement with the results in Table~\ref{tab:gainslosses_energy} for the $\vec{E} \cdot \vec{J}_{\rm e}$ integrals within the plasma sheath region around the waverider hence confirming the validity of our scaling law. Although the net electron cooling originating from the sheath everywhere in the domain can be observed to be approximately the same whether the GY or SE model is used, the \emph{sheath-induced electron cooling per electron within the plasma} is approximately 2 times larger  when using the GY model because the plasma density obtained with the GY model is about 2 times less (see electron density contours in Fig.~\ref{fig:Ne_contours_waverider}).  Therefore, this explains why the larger ion mobility of the GY model leads to lower electron temperatures. Indeed, the leading mechanism affecting electron temperature is shown to be electron cooling {within the sheath}, which is a function of ion mobility rather than electron mobility, and the ion mobility of the GY model is considerably higher than the one of the SE model.

Even small changes in electron temperature due to the ion mobility model can lead to significant plasma density changes because of the high sensitivity of three-body electron-ion recombination rates to electron temperature. In Earth-entry flows at less than Mach 25, most of the electron losses are generally due to 2-body recombination. But this is not the case when the electron temperature is much lower than the gas temperature. Indeed, the reaction rates for 3-body recombination scale with $T_{\rm e}^{-4.5}$  and thus increase by orders of magnitude when the electron temperature is low. For the case under consideration here, most of the electron losses due to chemical reactions are in fact due to three-body recombination and are hence very sensitive to a change in electron temperature. Because of the considerably larger amount of electron cooling per electron induced by the GY ion mobility model, the electron temperature drops by 10--20\% throughout the domain. This, in turn, leads to the electron-ion recombination rates increasing by 1.5--2.3 times, thus leading to an additional decrease in plasma density.

\subsection{Effect of Ion Mobility on Electron Cooling in Quasi-neutral Regions}

\begin{table}[t]
\center\fontsizetable
  \begin{threeparttable}
\tablecaption{Electron cooling integrals in W/m within the quasi-neutral region of the hypersonic waverider baseline case.}
\renewcommand{\arraystretch}{1.4}
\begin{tabular*}{\columnwidth}{@{}l@{\extracolsep{\fill}}rr@{}}
\toprule
Electron energy gain-loss integral  & GY model   & SE model \\
\midrule
$\int_V  ({k_{\rm B} T_{\rm e}  }  \vec{V} 
\cdot \vec{\nabla} N_{\rm e}) \text{d}V$   & $0.00328$ & $0.00170$\\
$\int_{V}  - \mu_{\rm i} \frac{k_{\rm B}^2 T_{\rm e}   \left( T+T_{\rm e} \right)}{|C_{\rm e}| N_{\rm e}} \left|\vec{\nabla} N_{\rm e}\right|^2 \text{d}V$   & $-0.00415$ &$-0.00331$\\
$\int_V {k_{\rm B} T_{\rm e}  }  \vec{V} 
\cdot \vec{\nabla} N_{\rm e}
- \mu_{\rm i} \frac{k_{\rm B}^2 T_{\rm e}   \left( T+T_{\rm e} \right)}{|C_{\rm e}| N_{\rm e}} \left|\vec{\nabla} N_{\rm e}\right|^2 {\rm d}V
$   & $-0.00088$ &$-0.00161$\\
$\int_V  (\vec{E}\cdot\vec{J}_{\rm e}) \text{d}V$   & $-0.00161$ & $-0.00164$\\
\bottomrule
\end{tabular*}
\label{tab:edotje_neutral_integrals}
\end{threeparttable}
\end{table}

Not only does ion mobility affect electron cooling in the sheaths, it also does so within the quasi-neutral regions where ambipolar diffusion is present. We can easily show this by first noting that the electron velocity in the quasi-neutral regions becomes:
\begin{equation}
 \vec{V}_{\rm e} \approx  \vec{V} - \underbrace{\frac{\mu_{\rm i} k_{\rm B} T }{ |C_{\rm i}|}\left( 1+\frac{T_{\rm e}}{T} \right)}_{\begin{minipage}{2.1cm}\flushleft\footnotesize \scalefont{0.9} ambipolar diffusion coefficient\end{minipage}}  \frac{\vec{\nabla} N_{\rm e}}{N_{\rm e}}
\end{equation}
From the latter and noting that the electron current corresponds by definition to $\vec{J}_{\rm e}=\vec{V}_{\rm e} C_{\rm e} N_{\rm e}$ we obtain the following electron current density in the ambipolar diffusion regions:
\begin{equation}
\vec{J}_{\rm e} \approx C_{\rm e} N_{\rm e} \vec{V} 
+ \mu_{\rm i} k_{\rm B} \left( T+T_{\rm e} \right) \vec{\nabla} N_{\rm e}
\end{equation}
In finding the latter, we assumed that the dominant ion is singly charged  (i.e., $C_{\rm e}=-|C_{\rm i}|$). Now note that when there is no applied external field, the electric field within an ambipolar diffusion process is such that the electron flux due to drift approaches closely the electron flux due to diffusion. Thus, a good approximation for such ``ambipolar electric field'' can be obtained from Eq.~(\ref{eqn:Vk}) by setting both the electron and the bulk velocities to zero and isolating the electric field. This yields:
\begin{equation}
\vec{E} \approx - \frac{k_{\rm B} T_{\rm e}}{|C_{\rm e}| N_{\rm e}} \vec{\nabla} N_{\rm e}
\end{equation}
We can now take the dot product between the latter and the former equations to find an approximation to the electron energy input term within ambipolar diffusion regions free of an externally applied electric field:
\begin{equation}
\vec{E} \cdot \vec{J}_{\rm e} 
\approx
{k_{\rm B} T_{\rm e}  }  \vec{V} 
\cdot \vec{\nabla} N_{\rm e}
- \mu_{\rm i} \frac{k_{\rm B}^2 T_{\rm e}   \left( T+T_{\rm e} \right)}{|C_{\rm e}| N_{\rm e}} \left|\vec{\nabla} N_{\rm e}\right|^2
\label{eqn:ambipolarelectroncooling}
 \end{equation}
As observed in Table \ref{tab:edotje_neutral_integrals}, the latter expression provides a reasonable approximation for the electron energy input term, $\vec{E}\cdot\vec{J}_{\rm e}$, within the quasi-neutral regions of the baseline waverider flowfield when the plasma is not subjected to an externally applied electric field.

The first term on the RHS of Eq.\ (\ref{eqn:ambipolarelectroncooling}) can contribute to either electron heating or cooling, depending on whether the bulk plasma velocity is aligned with  or opposed to  the electron density gradient, respectively. This explains why significant electron heating due to $\vec{E}\cdot\vec{J}_{\rm e}$ occurs only between the bow shock ahead of the nose and the peak electron density (see Fig.\ \ref{fig:EdotJe_contours_waverider}). However, it is important to note that this term is only significant when the plasma bulk velocity is large, as in the case of the waverider problem. For other cases involving low-speed plasma flows or plasmas at rest, it would become negligible.

The second term on the RHS of Eq.\ (\ref{eqn:ambipolarelectroncooling}) is always negative and thus contributes solely to electron cooling. Interestingly, this effect does not depend on electron mobility but rather on \emph{ion} mobility. Since the first term is only significant for high-speed flows and the second term is always negative, it follows that, for low-speed flows or plasmas at rest without external electric fields, the electron energy source term, $\vec{E}\cdot\vec{J}_{\rm e}$, contributes only to electron cooling and is proportional to ion mobility. {However, for the hypersonic flow under consideration, the electron heating which does not depend on ion mobility is considerable and approaches in magnitude electron cooling. This is why we do not observe an impact of ion mobility on the $\vec{E}\cdot \vec{J}_{\rm e}$ term in the quasi-neutral regions as significant as in the non-neutral sheaths. }

\section{RAM-C-II Re-entry Vehicle Results}

\begin{figure}[!h]
     \centering
     \includegraphics[width=1.0\columnwidth]{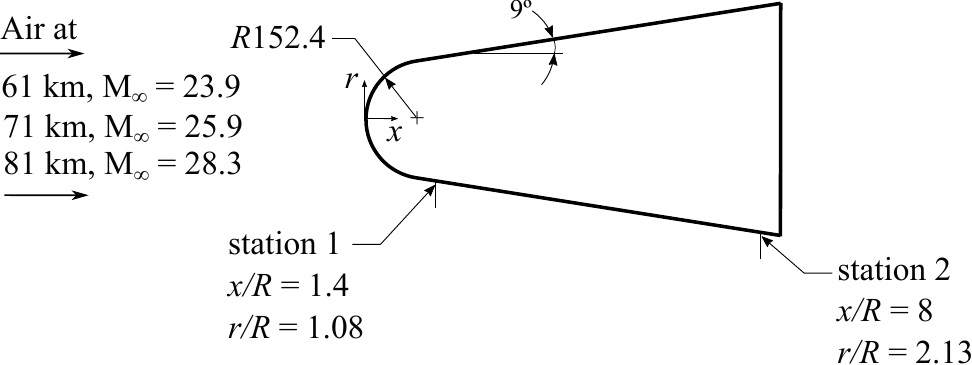}
     \figurecaption{Problem setup for the RAM-C-II case; dimensions in mm.}
     \label{fig:schematic_RAMCII}
\end{figure}

The second set of cases considered to assess the impact of ion mobility on plasma density and temperature is the re-entry flight test from the early 1970s known as RAM-C-II with the data outlined in \cite{nasa:1970:akey}. The geometry  consists of a 1.2-meter axisymmetric long body with a blunt leading-edge radius of 15~cm, followed by a 9-degree half angle cone, as shown in Fig.~\ref{fig:schematic_RAMCII}. Microwave reflectometers measured the peak electron number density near the stagnation region and along the truncated cone. Test data were obtained at a constant speed of 7650 m/s, at altitudes of 61~km, 71~km and 81~km. 
\begin{figure*}[!h]
     \centering
     \subfigure[61 km]{\includegraphics[width=0.27\textwidth]{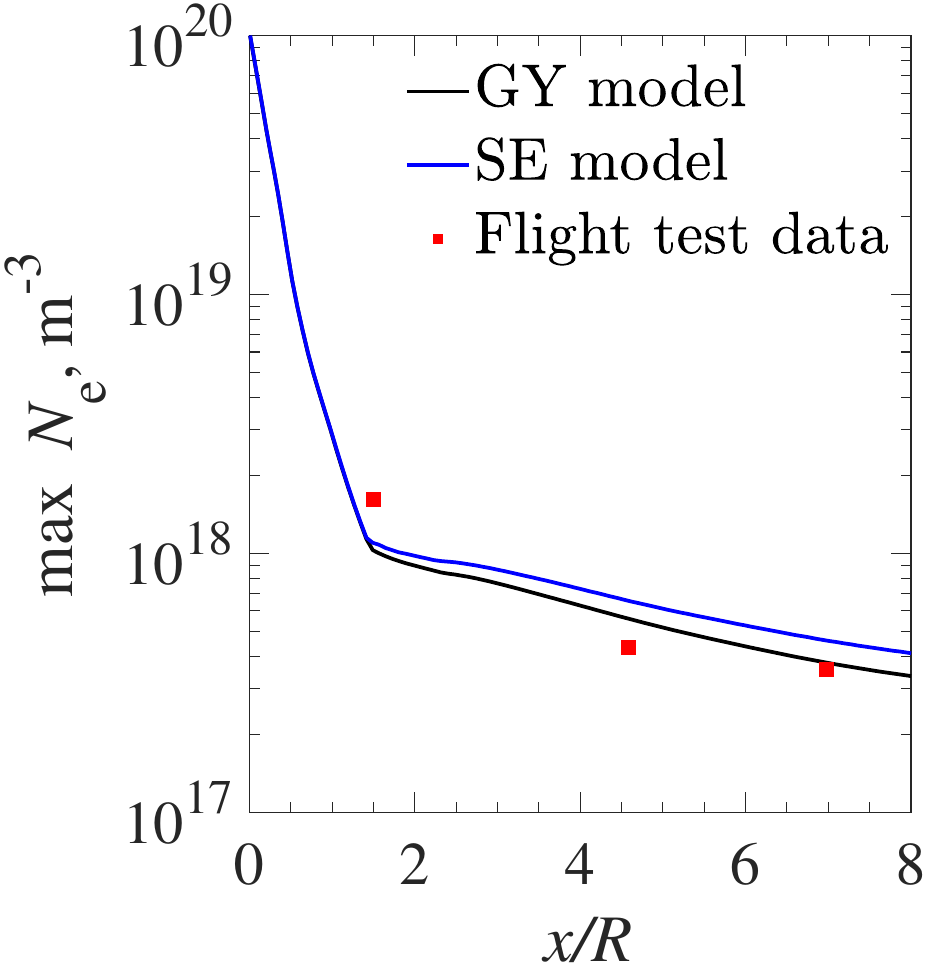}}
     ~~~~~~\subfigure[71 km]{\includegraphics[width=0.27\textwidth]{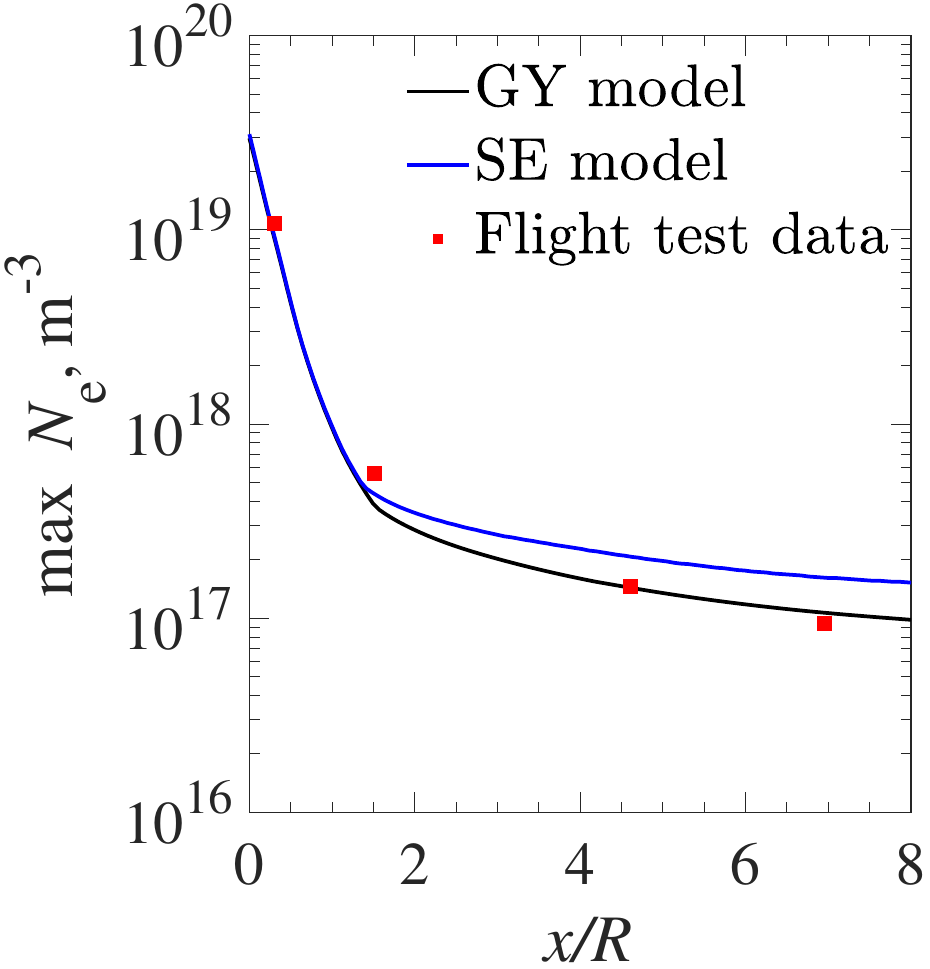}}
     ~~~~~~\subfigure[81 km]{\includegraphics[width=0.27\textwidth]{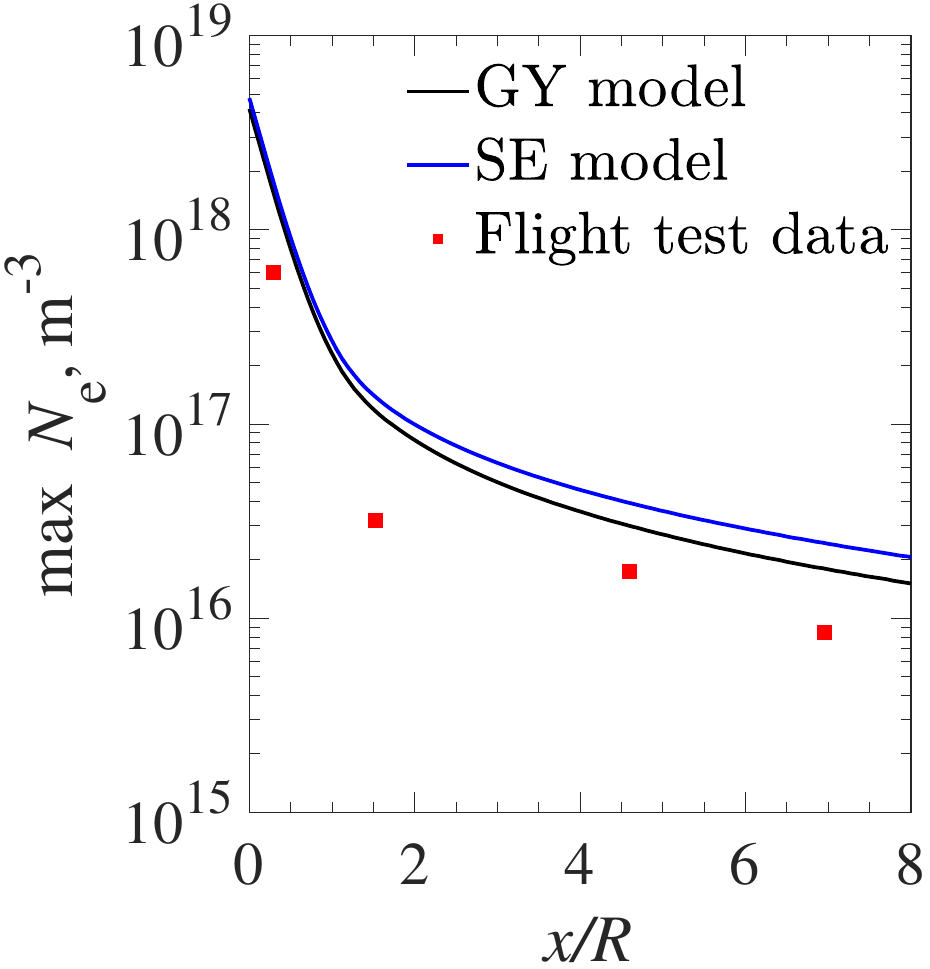}}

     \figurecaption{Maximum electron number density along the RAM-C-II body at (a) 61\ km altitude, (b) 71\ km altitude and (c) 81\ km altitude.}
     \label{fig:code_validation_ramcii}
\end{figure*}

A comparison of CFDWARP results with flight test data on the basis of maximum electron number density along the RAM-C-II axis is presented in Fig.~\ref{fig:code_validation_ramcii}, showing good agreement at both 61 km and 71 km and fair agreement at 81 km. The discrepancy between the numerical results and the flight test data at 81 km may be due to the flow being in the slip regime ($0.001\lesssim {\rm Kn} \lesssim 0.1$)  or due to some uncertainty about the freestream density and temperature at this altitude.  

\begin{figure*}[!h]
     \centering
     \subfigure[61 km]{\includegraphics[width=0.27\textwidth]{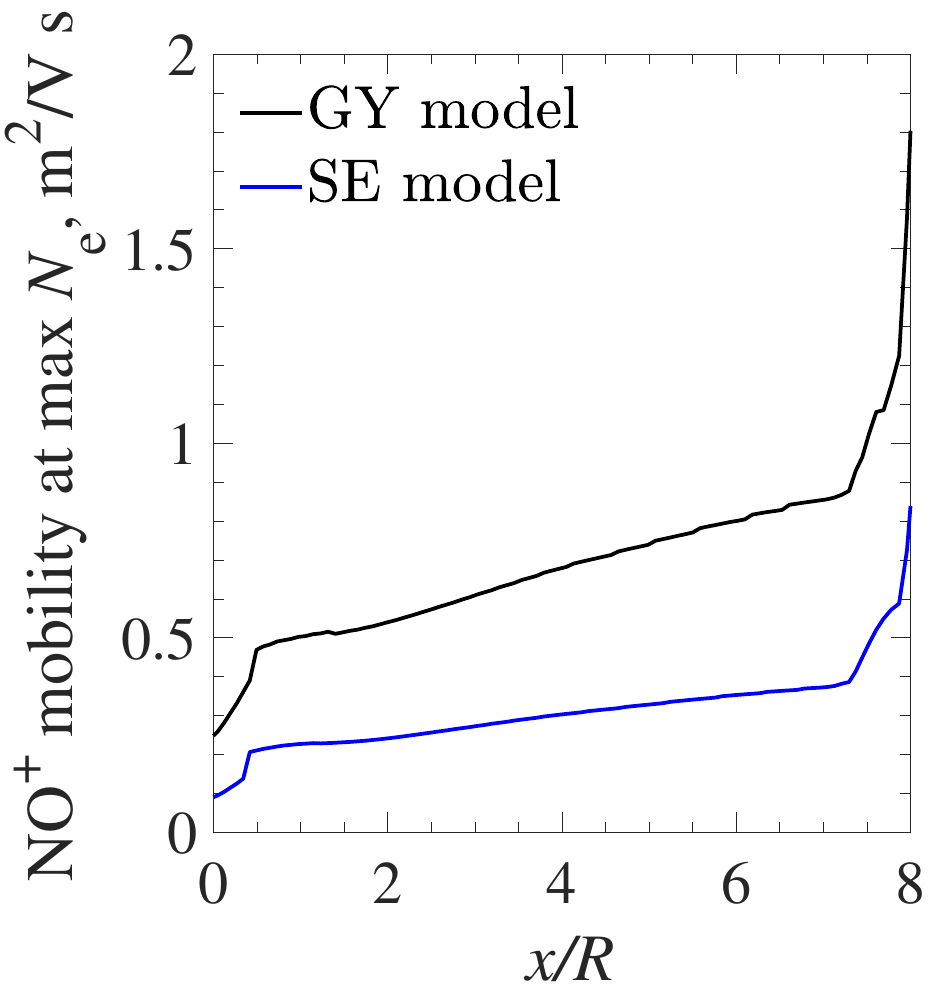}}
     ~~~~~~~\subfigure[71 km]{\includegraphics[width=0.27\textwidth]{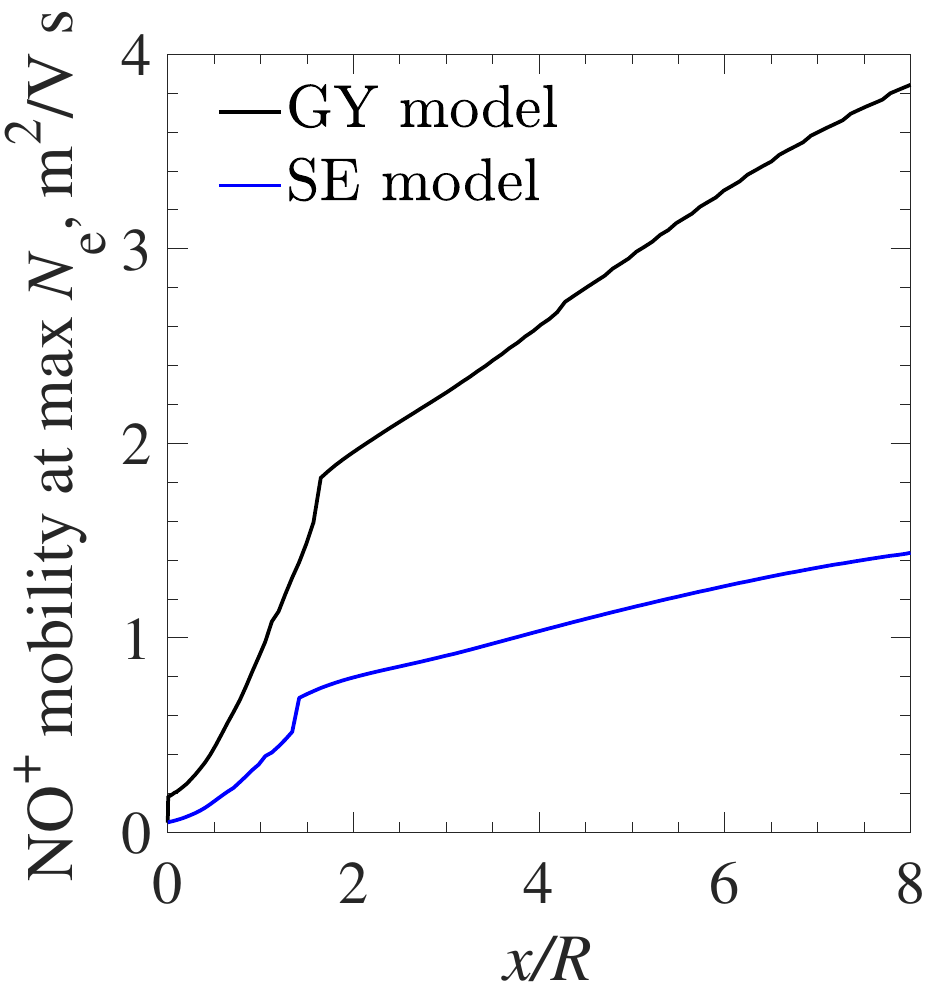}}
     ~~~~~~~\subfigure[81 km]{\includegraphics[width=0.27\textwidth]{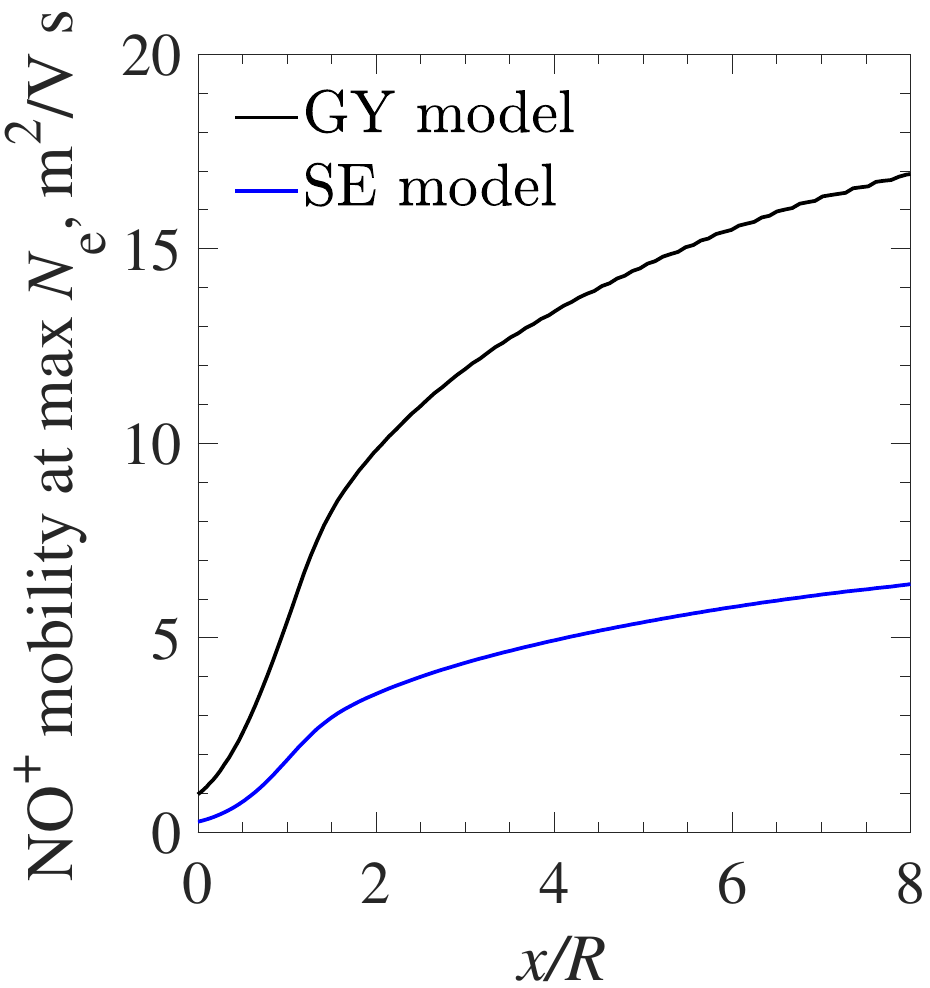}}
     \figurecaption{Dominant ion $\rm NO^+$ mobility along the RAM-C-II axis at the location of peak electron density for altitudes of  (a) 61 km, (b) 71 km, and (c) 81 km.}
     \label{fig:mobility_ramcii}
\end{figure*}

At all altitudes, we observe an impact of the ion mobility model on the maximum plasma density along the RAM-C-II body.   What is peculiar thus is that the biggest difference observed between the two models (about 60\%) occurs at the middle altitude of 71~km altitude and is significantly less (30\% or so) at either lower or higher altitudes. This can not be attributed simply to an effect of the  altitude on the ion mobility. Indeed, as can be seen from Fig.~\ref{fig:mobility_ramcii}, the difference between the ion mobilities of the two models is as much at the highest altitude than at the 71~km altitude.

\begin{figure*}[!h]
     \centering
     \subfigure[61 km]{\includegraphics[width=0.28\textwidth]{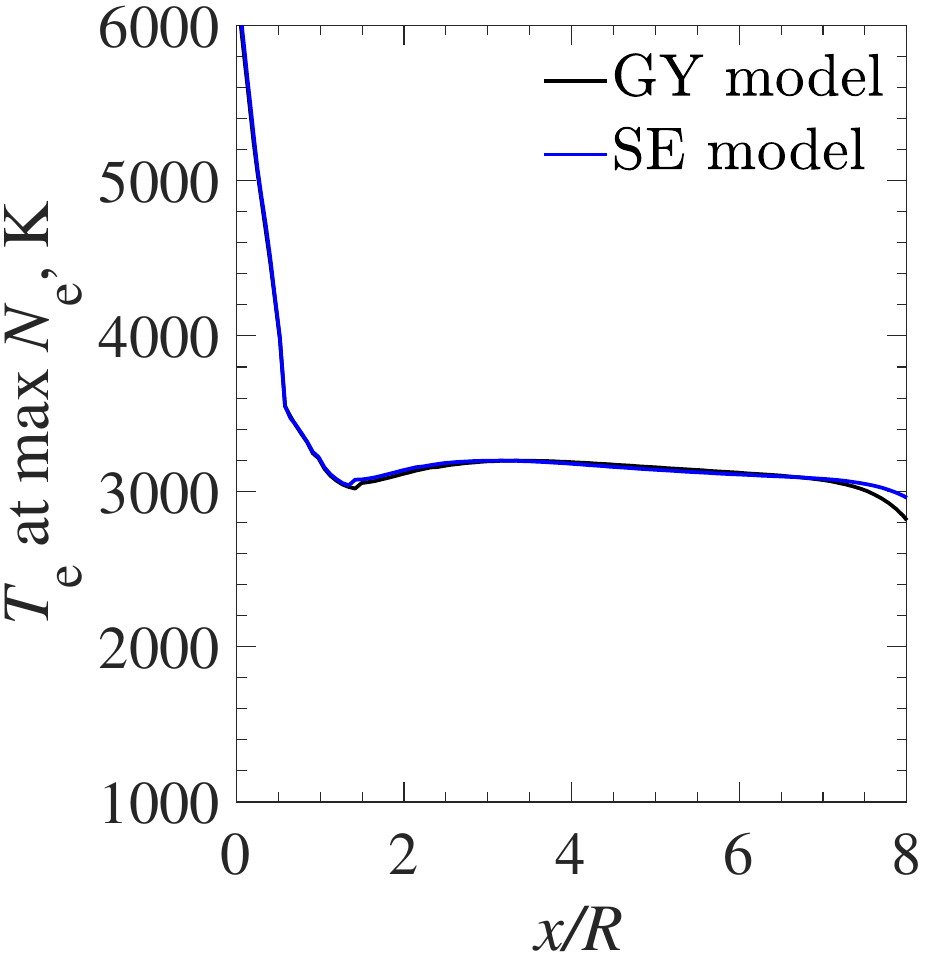}}
     ~~~~~~\subfigure[71 km]{\includegraphics[width=0.28\textwidth]{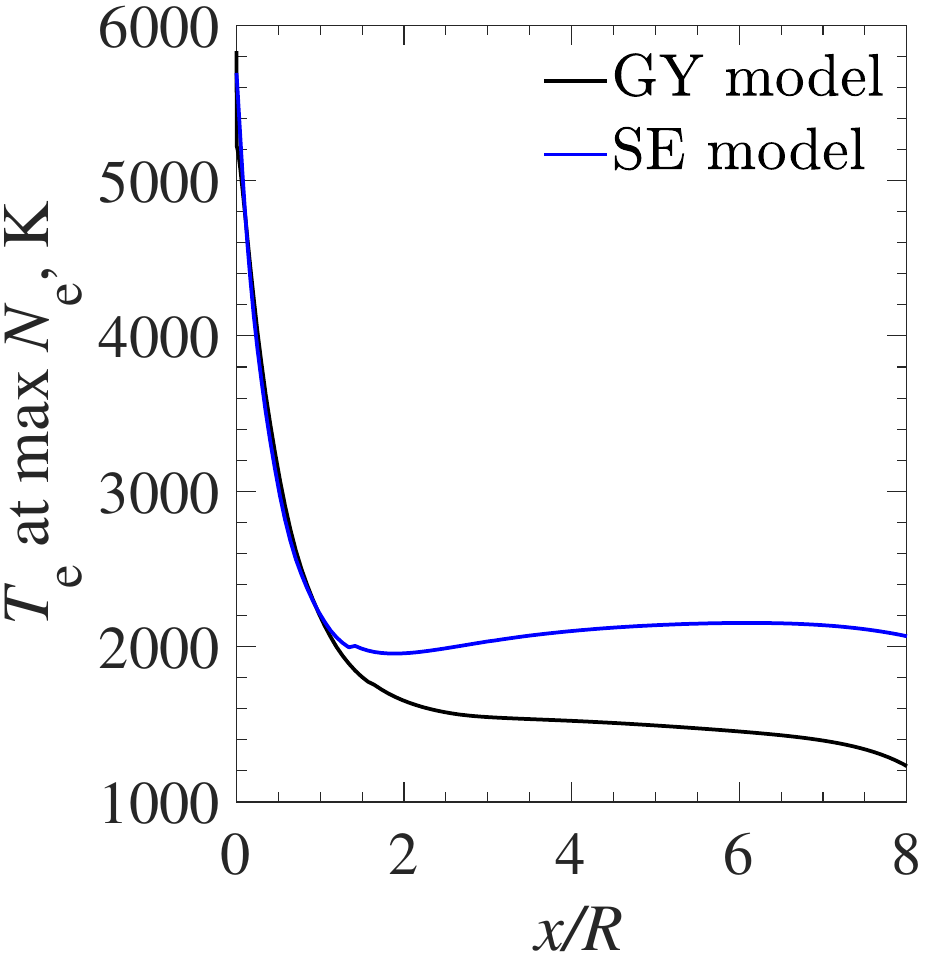}}
     ~~~~~~\subfigure[81 km]{\includegraphics[width=0.28\textwidth]{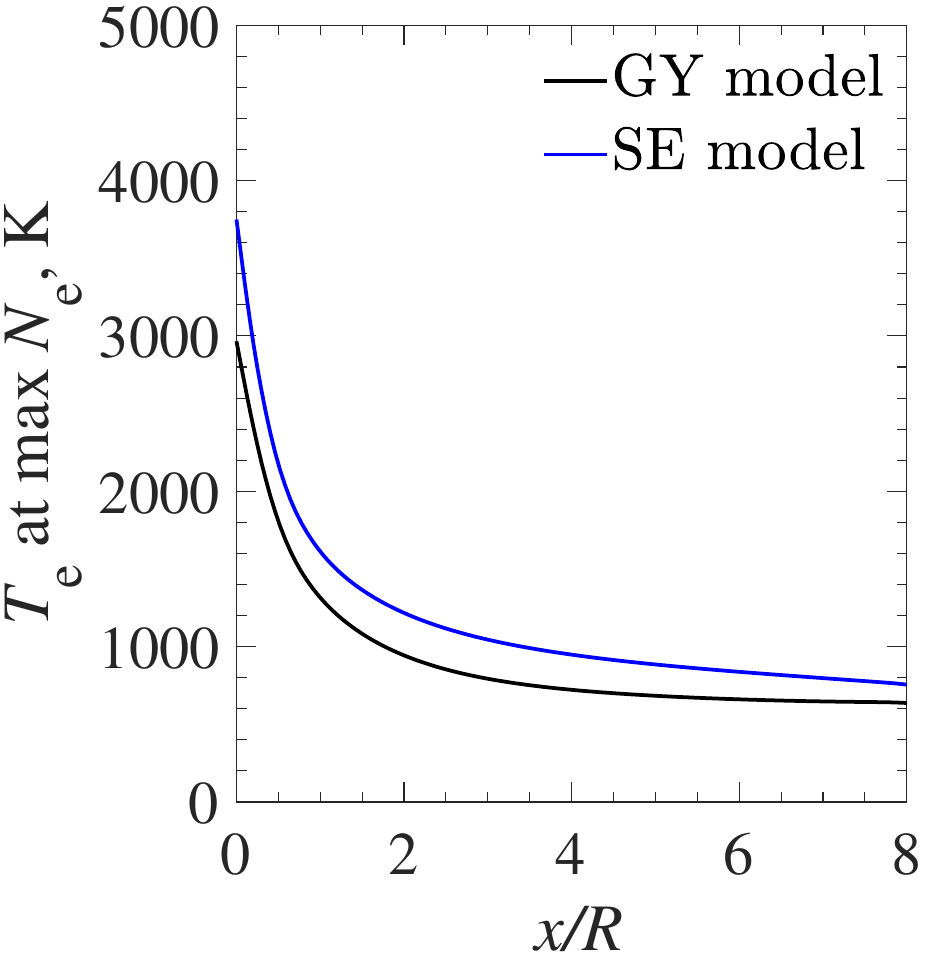}}
     \figurecaption{Electron temperature along the RAM-C-II axis at the location of maximum electron density at an altitude of (a) 61 km, (b) 71 km, and (c) 81 km.}
     \label{fig:Te_ramcii}
\end{figure*}
\begin{table*}[!t]
\center\fontsizetable
  \begin{threeparttable}
\renewcommand{\arraystretch}{1.3}
\tablecaption{Ratio of plasma density at station 2 and plasma density at station 1 for specific physical phenomena for the RAM-C-II cases.}
\fontsizetable
\begin{tabular*}{\textwidth}{@{}l@{\extracolsep{\fill}}cccccc@{}}
\toprule
~  & \multicolumn{2}{c}{61 km}  & \multicolumn{2}{c}{71 km}  & \multicolumn{2}{c}{81 km}\\
\cmidrule(lr){2-3}\cmidrule(lr){4-5}\cmidrule(lr){6-7}
Physical phenomena  & GY model   & SE model  & GY model    & SE model & GY model   & SE model  \\
\midrule
Ambipolar diffusion estimate, Eq.~(\ref{eqn:deltane_estimate}) & $0.72$  & $0.83$  & $0.52$ & $0.62$  & $0.49$ & $0.63$  \\
Chemical reactions (measured) & $0.92$  & $0.92$  & $0.87$ & $0.93$  & $0.72$ & $0.62$  \\
Axisymmetry effect estimate & $0.44$  & $0.44$  & $0.44$ & $0.44$  & $0.44$ & $0.44$  \\
Three above phenomena combined  & $0.29$  & $0.34$  & $0.20$ & $0.25$  & $0.15$ & $0.17$  \\
All measured from CFDWARP results  & $0.29$  & $0.36$  & $0.21$ & $0.33$  & $0.12$ & $0.14$  \\
\bottomrule
\end{tabular*}
\label{tab:Ne2_Ne1_altitudes}
\end{threeparttable}
\end{table*}

To understand better what phenomena are at play in reducing plasma density along the RAM-C-II body, we first derive a theoretical expression  for peak electron loss due to ambipolar diffusion along the truncated cone region between station 1 and 2 in Fig.~\ref{fig:schematic_RAMCII}. We here focus on the truncated cone region rather than the nose region because ion mobility is here seen not to affect significantly plasma density in the nose region. First, following \cite{jcp:2011:parent:2} we note that by defining the ambipolar diffusion coefficient as:
\begin{equation}
    D_{\rm a}=\frac{\mu_{\rm i} k_{\rm B} T}{C_{\rm i} } \left(1+\frac{T_{\rm e}}{T}\right)
\end{equation}
The electron mass conservation equation in ambipolar form neglecting chemical reactions becomes:
\begin{equation}
\frac{{\rm d} N_{\rm e}}{{\rm d} t} = -\frac{\partial}{\partial \chi}\left(D_{\rm a} \frac{\partial N_{\rm e}}{\partial \chi} \right)
\label{eqn:dNedt_RAMCII}
\end{equation}
where $\chi$ is a coordinate perpendicular to the surface and d/d$t$ refers to the substantial derivative. We can approximate the LHS by assuming the flow on the streamline travels at the constant speed $U$. Then, the LHS can be easily shown to become $U {\rm d} N_{\rm e}/{\rm d}s$ with $s$ a coordinate along the path from station 1 to 2. The latter can further be discretized to yield $U (N_{\rm e2}-N_{\rm e1})/d_{21}$ where $d_{21}$ is the distance between station 1 and 2 and $N_{\rm e1}$ is the peak electron number density at station 1 and $N_{\rm e2}$ is the peak electron number density at station 2. Also, noting that the electron density is orders of magnitude smaller at the surface than in the nearby plasma, we can approximate the spatial derivative of the electron density between the peak and the wall by the algebraic expression $\frac{1}{2}(N_{\rm e1}+N_{\rm e2})/ d_{\rm pw}$ with $d_{\rm pw}$ the distance between the peak plasma density and the wall.  Similarly, noting that the electron density at the shock is negligible compared to the one at the peak we can approximate the spatial derivative of the electron density between the shock and the peak by $-\frac{1}{2}(N_{\rm e1}+N_{\rm e2})/ d_{\rm ps}$ with $d_{\rm ps}$ the distance between the shock and the peak plasma density. Then, assuming small spatial variations in the ambipolar diffusion coefficient, we obtain the following discretized equation governing the change in peak electron density:
\begin{equation}
 \frac{U (N_{\rm e2}-N_{\rm e1})}{d_{21}} \approx -\frac{D_{\rm a}}{\frac{1}{2}(d_{\rm ps}+d_{\rm pw})}\left( -\frac{\frac{1}{2}(N_{\rm e1}+N_{\rm e2})}{d_{\rm ps}} -\frac{\frac{1}{2}(N_{\rm e1}+N_{\rm e2})}{d_{\rm pw}}\right)
\label{eqn:dNedt_RAMCII}
\end{equation}
We can further simplify the latter noting that $d_{\rm ps}\approx d_{\rm pw}$.  Then, isolating $N_{\rm e2}/N_{\rm e1}$, we arrive at the following expression for the electron density ratio due to ambipolar diffusion losses:
\begin{equation}
\dfrac{N_{\rm e2}}{N_{\rm e1}}\approx \frac{1-\eta}{1+\eta}
~~~~~~{\rm with}~~~~~\eta=\frac{D_{\rm a} d_{21}}{U d_{\rm pw}^2}=\dfrac{\mu_i k_B (T+T_{\rm e}) d_{21}}{|C_{\rm i}| d_{\rm pw}^2 U} 
\label{eqn:deltane_estimate}
\end{equation}
The latter expression provides a good estimate for the reduction in peak electron density due to ambipolar diffusion within the truncated cone region (between stations 1 and 2). It is emphasized that this takes into consideration the electron loss due to diffusion both towards the surface and towards the low plasma density region near the shock.

We assess the validity of our derived expression by comparing it with measured CFDWARP results for plasma density loss. First, we obtain a solution using CFDWARP and then analyze all properties within $\eta$, halfway between stations 1 and 2. It is crucial to include all physical phenomena that could affect plasma density in our prediction, such as ambipolar diffusion, axisymmetric flow, and chemical reactions. To estimate the reduction in plasma density due to axisymmetric effects, we perform a non-reacting simulation and measure the density at stations 1 and 2. Additionally, we estimate the impact of chemical reactions on plasma density by integrating the electron chemical source term, $W_{\rm e}$, along the streamline between stations 1 and 2. We combine these effects by multiplying the electron density ratios (i.e., the axisymmetric density ratio, the ambipolar density ratio, and the chemical reaction density ratio). This \emph{predicted} electron density ratio is then compared to the \emph{measured} electron density ratio from CFDWARP simulations of the RAM-C-II flowfield. As shown in the last two rows of Table \ref{tab:Ne2_Ne1_altitudes}, our theoretical predictions closely match the measured CFD results across all altitudes. The predictions also align with trends observed when switching between ion mobility models, giving confidence in the expression shown in Eq.~(\ref{eqn:deltane_estimate}).

Recall that our goal is to explain why ion mobility affects plasma density much more at  71 km than at lower or higher altitudes. Intuitively, we might expect ambipolar diffusion to become stronger at higher altitudes, as it is proportional to ion mobility, which increases at lower densities due to fewer ion-neutral collisions. Indeed, as shown in Fig.~\ref{fig:mobility_ramcii}, ion mobility increases several times for both ion mobility models across the altitudes considered. However, despite ion mobility increasing nearly fivefold between 71 km and 81 km for either the GY or SE model, this does not result in a greater impact of the ion mobility model on plasma peak density.

The smaller-than-expected impact of ion mobility on electron density peak erosion due to ambipolar diffusion is because the latter depends not only on ion mobility but also on electron temperature and the distance between the plasma peak density and the wall (see Eq. (\ref{eqn:dNedt_RAMCII})). We find that increases in ion mobility are consistently accompanied by (i) a decrease in electron temperature (see Fig. \ref{fig:Te_ramcii}) and (ii) a considerable increase in the distance between the wall and the peak electron density. Both effects significantly reduce the influence of ion mobility on ambipolar diffusion. 

On the other hand, the impact of ion mobility on electron temperature also affects chemical reaction rates, particularly two-body electron-ion recombination. Since two-body electron-ion recombination rates are inversely proportional to the square root of the electron temperature, a twofold decrease in electron temperature would be expected to increase recombination rates by about 40\%. This aligns well with the electron losses due to chemical reactions shown in Table \ref{tab:Ne2_Ne1_altitudes} and the observed effect of the ion mobility model on electron temperature in Fig.\ \ref{fig:Te_ramcii}. This is the main reason why the 71 km case shows a more pronounced impact of the ion mobility model on peak electron density compared to the other altitudes.

\section{Conclusions}

Ion mobility plays a pivotal role in determining plasma density around hypersonic waveriders, particularly under conditions of low flight dynamic pressure and high Mach numbers. Under these conditions, a change in the ion mobility model can result in more than a twofold variation in plasma density. Notably, the sensitivity of plasma density to different ion mobility models is on par with its sensitivity to variations in chemical models and electron energy transport models. This highlights the importance of accurately modeling ion mobility in predicting plasma behavior in hypersonic flows.

One primary mechanism by which ion mobility influences electron density is through ambipolar diffusion. In the quasi-neutral plasma surrounding hypersonic vehicles, electron diffusion is restrained by the ambipolar electric field, resulting in an effective electron diffusion coefficient that is directly proportional to ion mobility. In plasma flows where electron loss via chemical reactions is minimal compared to losses through surface catalysis -- such as in high-altitude, low-dynamic-pressure environments -- electron density is significantly impacted by ambipolar diffusion and, consequently, by ion mobility.

Additionally, ion mobility affects electron density through electron cooling. In this work, we derive novel scaling laws that demonstrate electron cooling scales proportionally with ion mobility in both the quasi-neutral and sheath regions.  Due to fast electron thermal diffusion, the ion-mobility-dependent cooling within the sheath contributes to a reduction in electron temperature throughout the plasma bulk, resulting in nearly a twofold increase in electron-ion recombination rates within the whole plasma layer.

These findings underscore the significant impact of ion mobility on both plasma density and electron temperature in hypersonic flows. Understanding and accurately modeling ion mobility are essential for predicting hypersonic plasma behavior, with implications for optimizing MHD technologies and for mitigating or harnessing
the interference caused by plasma layers on electromagnetic waves.

{Based on a continuum model, this study has limitations in simulating low-density flows near the slip regime and in simulating electron cooling in Debye sheaths. Ion mobility modeling also requires improvement, as one model showed significant errors at high temperatures and another at low temperatures. Other challenges include uncertainties in the secondary electron emission (SEE) coefficient at dielectric boundaries and incomplete understanding of electron heating rates due to electron-vibrational coupling. Enhancing the modeling of continuum processes, SEE, electron heating, and ion mobility is needed for improved accuracy in future studies.}

\footnotesize
\bibliography{all}
\bibliographystyle{plainnatmod}

\end{document}